\documentclass[onecolumn]{IEEEtran}

\usepackage{ragged2e}  
\usepackage{booktabs, makecell, 
	tabularx}  
\newcolumntype{L}[1]{>{\RaggedRight\hspace{0pt}%
		\hsize=#1\hsize}X} 
\usepackage{siunitx} 
\usepackage{float}
\usepackage{threeparttable}
\PassOptionsToPackage{hyphens}{url}\usepackage{hyperref}
\usepackage[english]{babel}
\usepackage[autostyle]{csquotes}
\usepackage{comment}
\usepackage{subcaption}
\usepackage{mwe}
\usepackage{ragged2e}
\usepackage[autostyle]{csquotes}
\usepackage[utf8]{inputenc}
\usepackage{pgfplots}
\usepackage{svg}       
\usepackage{amsmath}
\usepackage[ruled,linesnumbered]{algorithm2e}

\usepackage[T1]{fontenc}
\DeclareMathVersion{concmath}
\SetMathAlphabet{\mathrm}{concmath}{\encodingdefault}{ccr}{m}{n}

\newlength{\mylength}
\makeatletter
\newcommand{\mycfs}[1]{%
	\normalsize
	\@defaultunits\mylength=#1pt\relax\@nnil
	\edef\@tempa{{\strip@pt\mylength}}%
	\ifx\protect\@typeset@protect
	\edef\@currsize{\noexpand\mycfs\@tempa}
	\fi
	\mylength=1.2\mylength
	\edef\@tempa{\@tempa{\strip@pt\mylength}}%
	\expandafter\fontsize\@tempa
	\selectfont
}

\usepackage{cite}
\usepackage{amsmath,amssymb,amsfonts}
\usepackage{algorithmic}
\usepackage{graphicx}
\usepackage{textcomp}
\usepackage{xcolor}
\def\BibTeX{{\rm B\kern-.05em{\sc i\kern-.025em b}\kern-.08em
		T\kern-.1667em\lower.7ex\hbox{E}\kern-.125emX}}
\IEEEoverridecommandlockouts

\usepackage{xcolor}
\hypersetup{
	colorlinks,
	linkcolor={red!50!black},
	citecolor={blue!50!black},
	urlcolor={blue!80!black}
}

\ifCLASSINFOpdf
\else
\fi

\begin{document}
	\title{Sharding Distributed Databases:\\A Critical Review*\\
		{\footnotesize \textsuperscript{\textbf{*}}\textbf{Note:} \textit{This article is mostly Chapter 3 of the PhD dissertation titled: ``Novel Fault-Tolerant, Self-Configurable, Scalable, Secure, Decentralized, and}} \\
		\vspace{-0.7em}
		{\footnotesize \textit{High-Performance Distributed Database Replication Architecture Using Innovative Sharding to Enable the Use of BFT Consensus Mechanisms in Very}} \\
		\vspace{-0.7em}
		{\footnotesize \textit{Large-Scale Networks'', authored  by Siamak Solat. The complete dissertation is accessible via the following link on ResearchGate:} \href{https://www.researchgate.net/publication/379148513_Novel_Fault-Tolerant_Self-Configurable_Scalable_Secure_Decentralized_and_High-Performance_Distributed_Database_Replication_Architecture_Using_Innovative_Sharding_to_Enable_the_Use_of_BFT_Consensus_Mec}{PhD Dissertation}.}
	}
	
	\author{
		\IEEEauthorblockN{Siamak Solat}\\
		\IEEEauthorblockA{Université Paris Cité (formerly known as Université de Paris) \\
			Paris, France \\
			siamak.solat@etu.u-paris.fr}
	}
	
	\maketitle
	
	\begin{abstract}
		\noindent
		This article examines the significant challenges encountered in implementing sharding within distributed replication systems.
		It identifies the impediments of achieving consensus among large participant sets, leading to scalability, throughput, and performance limitations.
		These issues primarily arise due to the message complexity inherent in consensus mechanisms.
		In response, we investigate the potential of sharding to mitigate these challenges, analyzing current implementations within distributed replication systems.
		Additionally, we offer a comprehensive review of replication systems, encompassing both classical distributed databases as well as Distributed Ledger Technologies (DLTs) employing sharding techniques.
		Through this analysis, the article aims to provide insights into addressing the scalability and performance concerns in distributed replication systems.
	\end{abstract}
	
	
	\IEEEpeerreviewmaketitle
	
	\section{Introduction}
	\label{Introduction}
\begin{figure}
		\centering
		\includegraphics[width=0.9\linewidth]{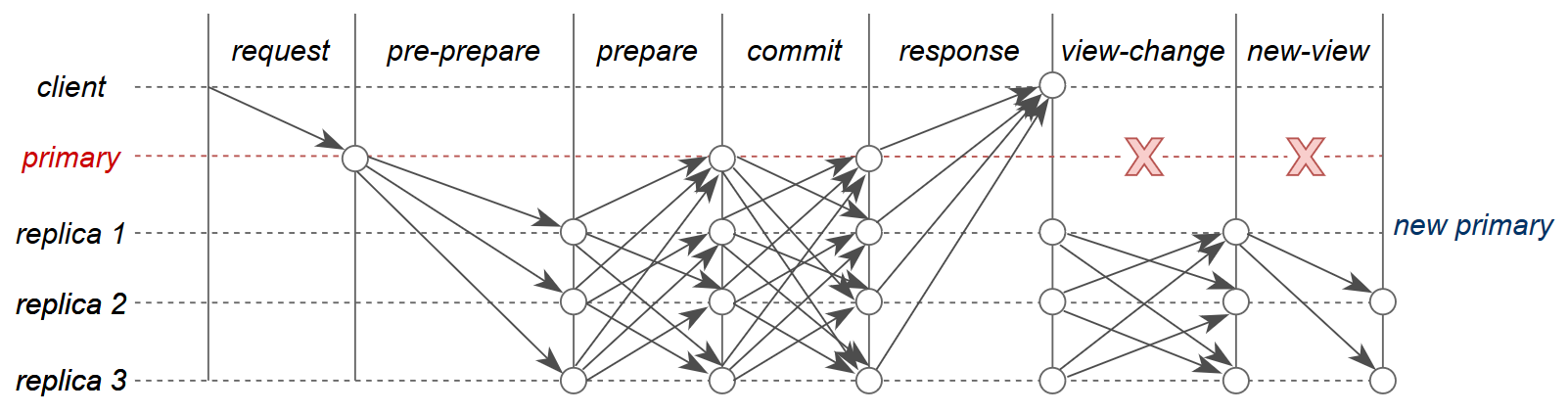}
		\caption{\footnotesize The PBFT consensus message complexity where a primary node fails and a change-view with additional message exchange is required, so that for $f$ leader failures the message complexity increases to $O(f.n^3)$.}
		\label{consensus_message_complexity}
	\end{figure}

\begin{figure}
		\centering
		\includegraphics[width=0.5\linewidth]{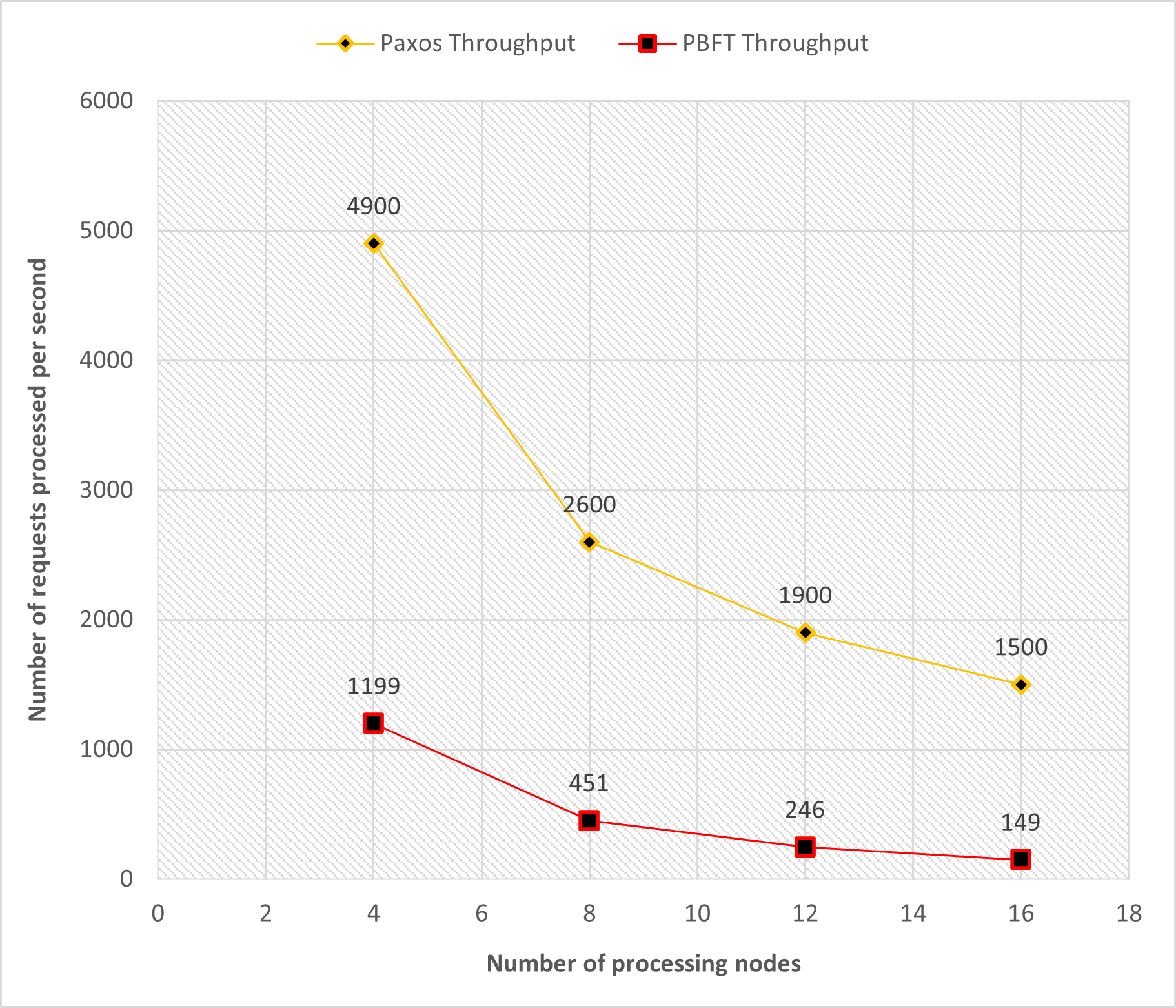}
		\caption{\footnotesize Throughput of a network that uses Paxos or PBFT consensus decreases drastically, as the number of nodes increases \cite{blockchain_consensus_protocols}.}
		\label{low_throughput_paxos_pbft}
	\end{figure}

	\noindent
	\IEEEPARstart{M}{ost} existing Byzantine fault-tolerant algorithms are very slow and are not designed for large sets of participants trying to reach a consensus.
	Hence, distributed replication systems that use consensus mechanisms to process clients' requests have major limitations and problems in scalability, throughput, and performance.
	The scalability problem means that system performance and throughput slows down as the number of network nodes increases.
	Such problems are mainly due to the message complexity of the consensus algorithms.
	For example, the message complexity of PBFT \cite{PBFT} is $ O(n^2) $ and that of Paxos \cite{Paxos} and Raft \cite{Raft} is $ O(n) $.
	While PBFT is both crash and Byzantine fault-tolerant, Paxos and Raft are only crash fault-tolerant.
	These message complexities can even be exacerbated with the presence of faulty nodes, when the faulty leader/primary node must be replaced through a view-change process. In a consensus mechanism, a view-change means switching to a new leader node. The view-change as an algorithm for choosing a new leader to collect information and propose it to processor nodes is the epicenter of a replication system \cite{Hotstuff}. For example, in the case of PBFT and Paxos, when the leader/primary node fails, the message complexity is exacerbated to $ O(n^4) $ and $ O(n^2) $, respectively \cite{Hotstuff}. Figure \ref{consensus_message_complexity} depicts the number of required message exchanges between nodes in the PBFT consensus algorithm.
	In any case, as the number of nodes in the network increases, the average processing time of clients' requests also increases, which ultimately leads to a great limitation for scaling up the network. Figure \ref{low_throughput_paxos_pbft} depicts how the mean time to process clients' requests increases as the number of nodes in the network increases when using the Paxos or PBFT consensus mechanisms.\\
	
	\noindent Even by replacing classic consensus mechanisms with Proof-of-Work (PoW) on networks similar to Bitcoin \cite{Bitcoin} there are still limits to the scalability, performance, and throughput, as the throughput of the Bitcoin network is only about $\approx$ 7-10 transactions per second \cite{Zilliqa}.
	Albeit, in general, there is a controversy and a difference of opinion in recognizing PoW as a consensus because there is a belief that it does not have the required properties of a consensus mechanism \cite{pow_is_not_consensus_1,pow_is_not_consensus_2,pow_is_not_consensus_3,pow_is_not_consensus_4}.
	The problem of scalability becomes very important and crucial when the network is open or so-called permissionless, because no permission from any privileged entity is required to create processor nodes and participate in processing requests or to create client nodes and sending requests.
	This is why platforms such as Hyperledger \cite{Hyperledger} use permissioned networks to control the number of nodes by mandating the need for permission from some privileged entity to create processor nodes or send requests in order to limit the size of the network, otherwise by increasing the number of nodes the throughput of the network decreases dramatically due to time complexity of the consensus mechanism used for processing clients' requests.
	
	\section{Sharding at a Glance}
	\label{Sharding_at_a_glance}
	\noindent
	In a typical scenario, a single database system is well-equipped with storage and performance capabilities to handle the transaction processing needs of an enterprise. However, challenges arise when dealing with applications catering to millions or even billions of users, such as social media platforms or large-scale user-centric applications in major institutions like banks \cite{Database_system_concepts}.
	Imagine an organization that has developed an application relying on a centralized database. As the user base grows, the limitations of the centralized database become evident, struggling to meet the increasing storage and processing speed requirements. To address this, a commonly adopted strategy is the practice known as \enquote{sharding}. This involves the segmentation of data across multiple databases, with each database handling a subset of users. Sharding, fundamentally the distribution of data across multiple databases or machines, proves essential in achieving scalability and improved performance \cite{Database_system_concepts}.
	As the number of databases increases, the risk of potential failures also rises, resulting in a heightened probability of losing access to critical data. To mitigate this risk, replication becomes imperative to guarantee continued accessibility even in the face of failures. However, the management of these replicas introduces additional complexities, demanding careful attention to ensure their consistency and effectiveness \cite{Database_system_concepts}.\\
	
	\noindent
	Sharding is also used already in several blockchain-based systems in order to increase the scalability of the network.
	In a traditional blockchain system, all nodes on the network must process every transaction that occurs on the network. This means that as more transactions occur, the network can become congested and slow. Sharding solves this problem by breaking the network into smaller segments called shards.
	Each shard processes a subset of transactions rather than all transactions on the network. By distributing the workload across multiple shards, the network can handle more transactions per second and hence become more scalable.
	In a sharded replication system, each node is responsible for processing transactions only in its assigned shard. This reduces the computational requirements for each node and makes it easier for new nodes to join the network.
	Sharding is still an area of active research and development in distributed replication networks, but it has the potential to significantly improve the performance and scalability of blockchain networks.
	Considering the limitations and obstacles in consensus algorithms for scaling, one of the main reasons for such low throughput in replication systems that use consensus mechanisms is the serial processing approach, where each client request is processed by all processor nodes that leads to a significant decrease in system performance and throughput in a redundancy approach.
	In contrast to serial processing, sharding as a parallelization approach has already been implemented in several replication systems and has shown a notable capability and potential to improve performance and scalability, yet, current sharding techniques have several remarkable problems detailed in Section \ref{Sharding_challenges}. \\
	
	\noindent Replication systems such as Bitcoin, Ethereum, and Hyperledger that use consensus mechanisms---both classic and so-called \enquote{proof-of-x} techniques\footnote{Various methods of using blockchain systems to prove something in a way that is cryptographically verifiable \cite{proof-of-x}.}---to process requests and transactions have low throughput, performance, and scalability.
	The only reason why non-sharded Ethereum nodes can store the entire state (or the whole replication) is that Ethereum only processes around 15 transactions per second \cite{eth_tps}. Once a system processes thousands of transactions per second, the state will explode, since transactions do leave a trace on the state \cite{zilliqa_limitations}.
	A common way of dealing with clients' requests in such systems is serial processing approach, where all the requests are processed by all the processor nodes in the network and hence, by joining new nodes to the network the total request processing capacity of the system gradually will decrease. Networks that use classic consensus---either with linear or quadratic message complexity---to process clients' requests lead to increased processing coordination costs \cite{Nightshade}.
	On the other hand, networks that use PoW as an alternative to classic consensus algorithms face the same problem by increasing the computing power of the entire network.
	Even if the number of processor nodes gets limited by a centralized approach and using a privileged entity in a permissioned network, by increasing the rate of clients' requests, the processor nodes hardware performance is still limited, causing significant latency in response to the clients \cite{Nightshade}.
	One of the proposed solutions to this problem is using sharding technique by dividing the network into multiple smaller groups, each of which handles a part of clients' requests.
	Several protocols have been already proposed based on sharding technique. We describe briefly some of them in Section \ref{related_works_sharding}. \\
	
	\noindent The sharding technique can be divided into two general types \cite{Harmony_protocol}: \vspace{0.1cm}
	\begin{itemize}
		\item Processing Sharding \vspace{0.1cm}
		\item Storage/State Sharding \\
	\end{itemize}
	\vspace{-0.1cm}
	
	\noindent For example, Zilliqa \cite{Zilliqa} is not a state sharding protocol, as each node holds the entire stored replicated data state to be able to process transactions or clients' requests, while other solutions like Omniledger \cite{OmniLedger} and RapidChain \cite{Rapidchain} feature state sharding, where each shard holds a subset of the stored replicated data state. 
	In most of the cases, storage/state sharding typically brings us processing sharding as well. To the best of our knowledge, actually there is no protocol that uses storage sharding without processing sharding.\\
	
	\noindent The rest of this article is organized as follows:
	In Section \ref{Sharding_challenges}, we describe the most important challenges in the sharding of distributed replication systems, which are divided into the following four subsections: in subsection \ref{random} we explain why most current sharding protocols use a random assignment approach for allocating and distributing nodes between shards due to security reasons.
	In subsection \ref{process_tx_in_sharding} we detail how a transaction is processed in sharded DLTs, based on current sharding protocols.
	In subsection \ref{shared_state_problems} we describe how a shared ledger among shards imposes additional scalability limitations and security issues on the network.
	And in subsection \ref{undesirable_cross_shard_tx} we explain why cross-shard or inter-shard transactions are undesirable and more costly, due to the problems they cause, including atomicity failure and state transition challenges, along with a review of proposed solutions.
	And, in Section \ref{related_works_sharding}, we review some replication systems, including both classic distributed databases and DLTs, that utilize the sharding technique.

\section{Sharding Challenges}
\label{Sharding_challenges}
\noindent
Sharding, as a parallelization approach, has already been implemented in several distributed replication systems and has shown remarkable potential to improve performance and scalability; nevertheless, current sharding techniques face several challenges. We describe the most important of them below.

\subsection{Distributing Nodes Between Shards:}
\label{random}
\begin{figure}
		\centering
		\includegraphics[width=0.9\linewidth]{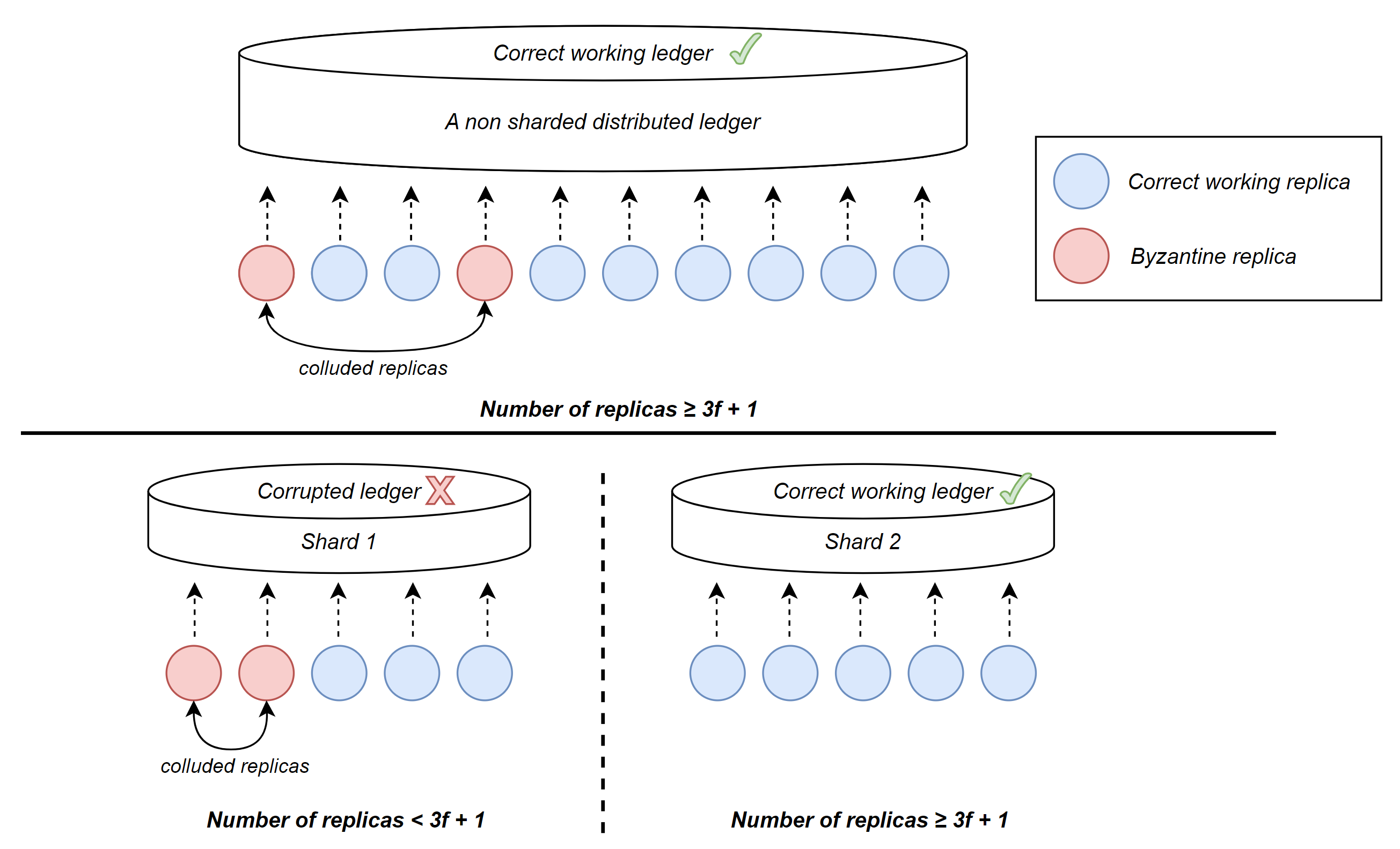}
		\caption{\footnotesize Most current sharding protocols use a random assignment approach for allocating and distributing nodes between shards due to security reasons.}
		\label{why_random-assignment}
	\end{figure}

\noindent
	Most current sharding protocols use a random assignment approach for allocating and distributing nodes between shards due to security reasons. We explain why this approach is employed in most sharding protocols using the following example: assume in a non-sharded replication system, there are in total 10 replicas, two of which are Byzantine and also know each other as the members of a cyber-attacker group, that is, they are \textit{colluded} replicas, as depicted in Figure \ref{why_random-assignment}. If the consensus is PBFT, the network can remain safe if the number of nodes, $n$, is greater than or equal to $3f + 1$, where $f$ is the number of Byzantine or faulty nodes.
	We then divide the network into two shards, in such a way that the replicas are permitted to choose which shard to assign. Obviously, two Byzantine replicas prefer to be the member of the same shard in order to be able to dominate that shard. Hence, in most sharding protocols, the assignment of replicas between shards is done in a random manner. This is to defeat the security problem because, in the case of using a random assignment approach, the probability that all the members of an attacker group or colluded replicas are assigned to the same shard is considerably reduced.
	
\subsection{Transactions Processing in Sharded DLTs:}	
\label{process_tx_in_sharding}
\noindent
	In this section, we detail how a transaction is processed in a sharded replication system, based on current sharding protocols. In state/storage sharding, processed transactions are stored in separate shards. With state sharding, each node has a shard it is assigned to in such a way that at any given moment of time, the state of the stored replicated data is split between shards in a way known to all nodes. In other words, all the nodes---and everything else that is stored in state---is split between the shards in some way known to all nodes. Each committee is assigned to a particular shard, which is responsible for every particular subset of the state. A transaction is affecting some nodes in the network, meaning that, if a client node $ n_{c\alpha} $ makes a transaction and sends some token to another client node $ n_{c\beta} $, both nodes are affected by this transaction, so that the token balance\footnote{The number of crypto-tokens that a node holds.} of $ n_{c\alpha} $ decreases, while that of $ n_{c\beta} $ increases.
	Each node is assigned to exactly one shard, and the transaction is processed by the committee, and only by the committee, that is responsible for the subset of the state that the transaction is affecting.
	If the nodes affected by the transaction are assigned to the same shard, the transaction is an intra-shard transaction. However, if each of the nodes is assigned to different shards, a cross-shard or inter-shard transaction occurs, and each participating shard has access to only part of the transaction data for processing. Consequently, processing an inter-shard transaction is more complex that an intra-shard transaction.

\subsection{Challenges With Shared Ledger Among Shards:}	
\label{shared_state_problems}
\noindent
	Some sharding-based protocols, such as PolkaDot \cite{Polkadot}, Cosmos-Hub \cite{Cosmos}, and Ethereum 2.0\footnote{Recently, Ethereum 2.0 underwent a change in the architecture of its sharding approach \cite{Ethereum_2_0_New_Sharding_2024}.} \cite{Ethereum_number_of_shards}, utilize a shared ledger among shards for various bookkeeping computations.
	These computations include coordinating and orchestrating shards, distributing nodes between shards, capturing snapshots of the latest state of shards, and managing cross-shard transactions.
	The workload on this shared ledger is proportional to the number of shards in the network \cite{Nightshade}.
	This shared ledger among shards goes by different names; for instance, it is called the \enquote{Beacon} chain in sharded Ethereum, the \enquote{Relay} chain in PolkaDot, or \enquote{Cosmos-Hub} in the Cosmos protocol. However, in this thesis, we refer to this ledger simply as the shared ledger among shards.
	Such a shared ledger imposes scalability limitations and additional security challenges on the system, which we describe in detail.\vspace{0.1cm}
	
	\begin{itemize}
		\item Scalability Issues Due to Shared Ledger Among Shards:\\
		\vspace{-3mm} \\
		The sharding is often advertised as a solution enabling linear scalability, meaning that as the number of nodes in the network increases, the throughput of the system increases at an almost linear rate \cite{Linear_Scalability_Sharding}. While it is in theory possible to design such a sharding protocol, any solution that uses the concept of a shared ledger among shards cannot achieve such scalability.
		Since a shared ledger among shards is itself a ledger with computation bounded by the computational capabilities of the nodes operating it, the number of shards is naturally limited \cite{Nightshade}.
		Figure \ref{shared-ledger} depicts this situation, in which a shared ledger holds snapshots of data from all shards and ultimately imposes significant constraints on network scalability.
		Although participating nodes in the shared ledger keep only the necessary data, growing the network and increasing the number of shards introduces scalability problems, resulting in limitations for the entire system.\\
		
		\begin{figure}
			\centering
			\includegraphics[width=0.7\textwidth]{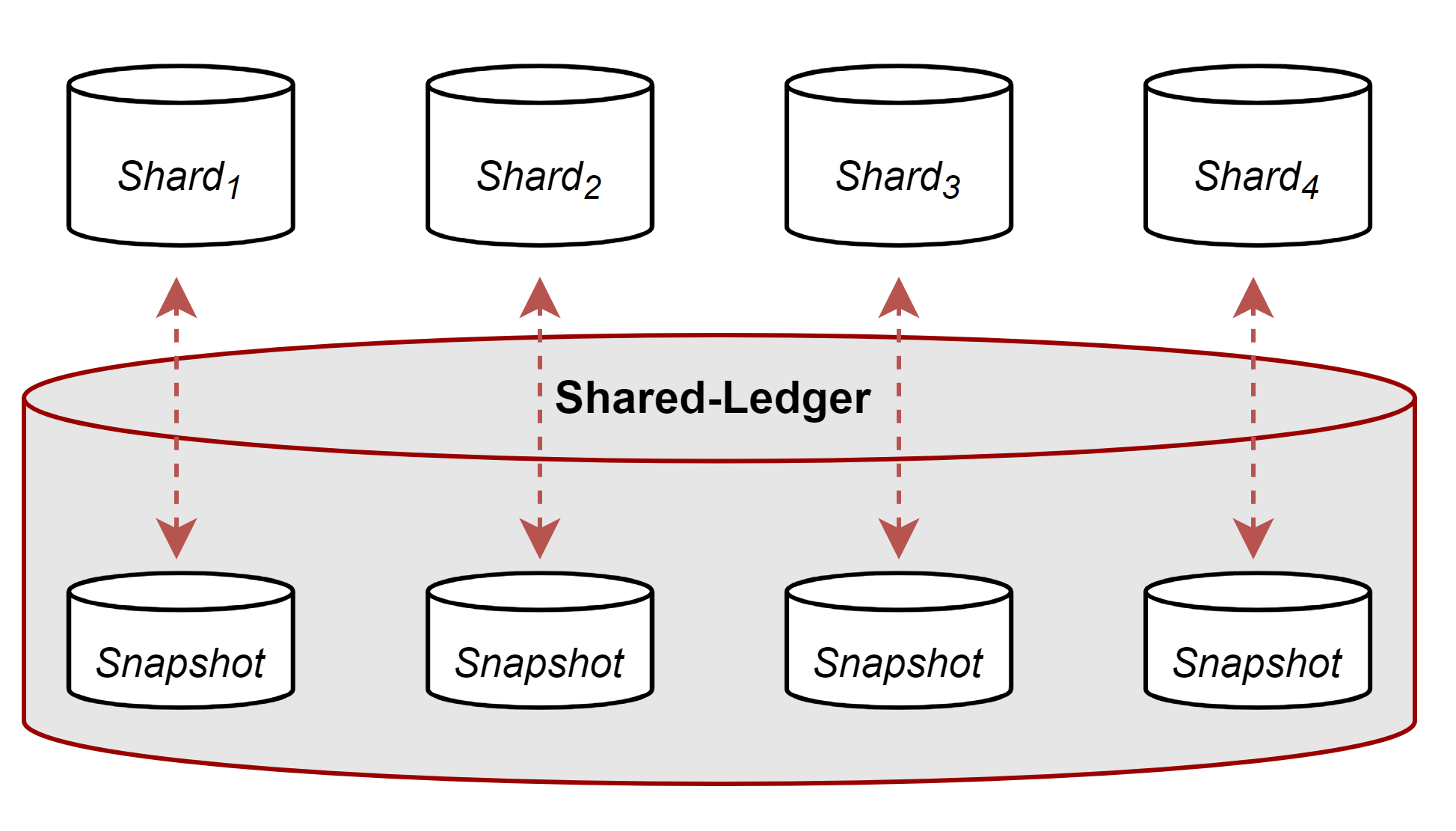}
			\caption{{\footnotesize A shared ledger among shards for various bookkeeping computations. These computations include coordinating and orchestrating shards, distributing nodes between shards, capturing snapshots of the latest state of shards, and managing cross-shard transactions. The workload on this shared ledger is proportional to the number of shards in the network \cite{Nightshade}.}}
			\label{shared-ledger}
		\end{figure}

		\item Security Issues Due to Shared Ledger Among Shards:\\
		\vspace{-3mm} \\
		In addition to the scalability challenge, if the nodes managing the shared ledger among shards turn Byzantine, the entire system becomes vulnerable. This is particularly problematic as critical tasks, such as node assignment between shards, are handled by this privileged shard. In essence, a compromised shared ledger, occupied by Byzantine nodes, has the potential to infect a significant portion of the system, putting it at risk of collapse. Given that a shared ledger is mission-critical, any flaw within it could compromise the integrity of the entire network \cite{beacon_chain_verification}.
	\end{itemize}

\subsection{Challenges With Cross-Shard Transactions:}
\label{undesirable_cross_shard_tx}
\noindent
	In a cross-shard transaction, each participating shard only has access to a portion of the transaction data for processing. Consequently, cross-shard transactions require costly inter-shard coordination to ensure state consistency, significantly limiting the system's performance. These circumstances make cross-shard transaction processing more complicated, complex, and therefore more expensive than intra-shard transactions. Before delving into the challenges associated with cross-shard transactions, it is essential to understand two general approaches for processing them.\\

		\noindent
	    - Synchronous Approach:
		In blockchain-based protocols, new states are equivalent to new blocks.
		In synchronous cross-shard transaction processing, new states (or blocks) containing state transitions related to a transaction are generated simultaneously.
		A cross-shard transaction visibly impacts a set of shards, as explained in Section \ref{process_tx_in_sharding}, and synchronous cross-shard transactions must be included at the same block height in all affected shards.
		To ensure a canonical order of execution, transactions within a block must be arranged based on their hash order.
		However, this approach necessitates a high level of coordination between shards, resulting in increased message complexity and prolonged time for creating new states (or blocks).
		This simultaneous block production, where all state transitions of a transaction occur at the same block height, means that blocks are generated as fast as the slowest shard.
		Consequently, this synchronous approach sacrifices the speed of intra-shard transactions for the communication overhead of cross-shard transactions. \\
		
		\noindent
	    - Asynchronous Approach:
		In asynchronous cross-shard transaction processing, each shard independently generates new states (or blocks).
		However, to ensure atomicity, a shard may need to lock a state transition, ensuring that the state is committed on another shard.
		This leads to higher transaction latency.
		In the asynchronous approach, blocks are generated more frequently than in the synchronous strategy.
		On the other hand, the processing time of a cross-shard transaction might be exacerbated.\\
	
	\noindent 
	To the best of our knowledge, while there are considerable discussions on the synchronous approach, as found in \cite{Merge_blocks, Synchronous_cross_shard}, there is still no notable sharded DLT protocol that employs a synchronous strategy for cross-shard transaction processing. Therefore, we place more emphasis on addressing the challenges associated with the asynchronous approach.\vspace{0.1cm}
	
	\begin{itemize}
		\item Atomicity Failure:\\
		\vspace{-3mm} \\
		In this section, we elaborate on the possibility of an atomicity failure occurring when using a sharded blockchain-based network during a cross-shard transaction.
		A financial transaction consists of two parts: credit (token-receiving) and debit (token-sending).
		Let's assume that each part is executed in a different shard as an inter-shard transaction.
		For a cross-shard transaction to maintain atomicity, it must either be committed in both shards or aborted uniformly across both.
		Failure to achieve this results in an atomicity failure.
		In simpler terms, the transaction must either be successfully committed, making all changes permanent and durable, or aborted, with all changes rolled back, undone, or discarded.
		This characteristic is known as atomicity, one of the ACID\footnote{Atomicity, Consistency, Isolation, Durability} transaction properties.
		In the asynchronous processing of a cross-shard transaction, if a fork occurs in one or both shards, and the chain of either the credit or debit parts becomes aborted as part of the forked chain while the other part remains in the canonical (main) chain, an atomicity failure occurs.
		This is because one part of the transaction is validated in a shard, while another part is abandoned in another shard.
		Figure \ref{atomicity_failure} depicts such a situation. \\
		
		\begin{figure*}
			\centering
			\includegraphics[width=1\textwidth]{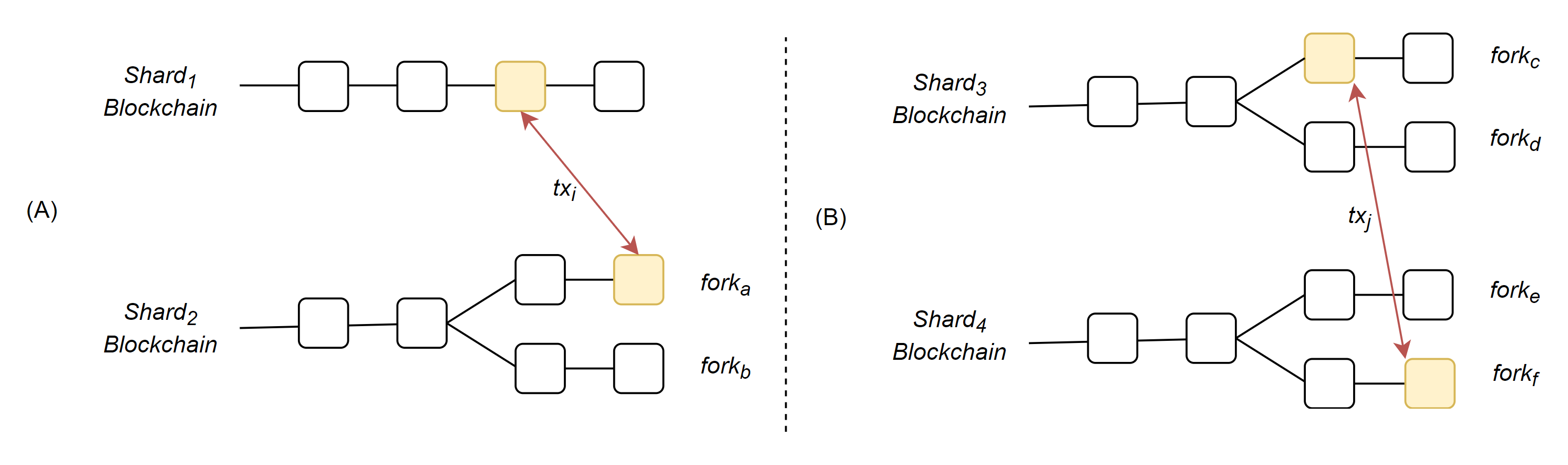}
			\caption{{\footnotesize (A): $\mathrm{tx_{i}}$ is a cross-shard transaction between $\mathrm{shard_{1}}$ and $\mathrm{shard_{2}}$, where a fork has occurred. If $\mathrm{fork_{a}}$ in $\mathrm{shard_{2}}$ as a part of transaction $\mathrm{tx_{i}}$ is in canonical/main chain, then $\mathrm{tx_{i}}$ gets finalized, otherwise an atomicity failure has occurred. (B): $\mathrm{tx_{j}}$ is a cross-shard transaction between $\mathrm{fork_{c}}$ in $\mathrm{shard_{3}}$ and $\mathrm{fork_{f}}$ in $\mathrm{shard_{4}}$. If both $\mathrm{fork_{c}}$ and $\mathrm{fork_{f}}$ are in canonical/main chain, then $\mathrm{tx_{j}}$ gets finalized. If both forks become abandoned, then $\mathrm{tx_{j}}$ becomes fully abandoned that is no conflict and the situation is fine. But if one of these forks becomes canonical/main chain, while another one is abandoned as a part of forked chain, then an atomicity failure has occurred.}}
			\label{atomicity_failure}
		\end{figure*}
		
		\item State Transition Challenge:\\
		\vspace{-3mm} \\
		Another challenge with cross-shard transactions is the potential for abusing this type of transaction to turn an invalid data transition into a valid one.
		Consider Figure \ref{state_transition-challenge_in-sharding}, in which $\mathrm{shard_{a}}$ is corrupted.
		A group of colluding Byzantine processing nodes creates an invalid block $a_{2}$, resulting in unexpected token minting on an account, denoted as $\psi$.
		Subsequently, the Byzantine nodes generate a valid block $a_{3}$ on top of the invalid block $a_{2}$.
		While the transactions in block $a_{3}$ are applied correctly, those in block $a_{2}$ are not executed properly.
		The Byzantine nodes then initiate a cross-shard transaction towards $\mathrm{shard_{b}}$, where all blocks are valid and in a correct state.
		They transfer the tokens from account $\psi$ to another account, $\xi$. 
		From this moment onward, the improperly created tokens reside on a fully valid ledger in $\mathrm{shard_{b}}$.\\
		
		\begin{figure}
			\centering
			\includegraphics[width=0.9\textwidth]{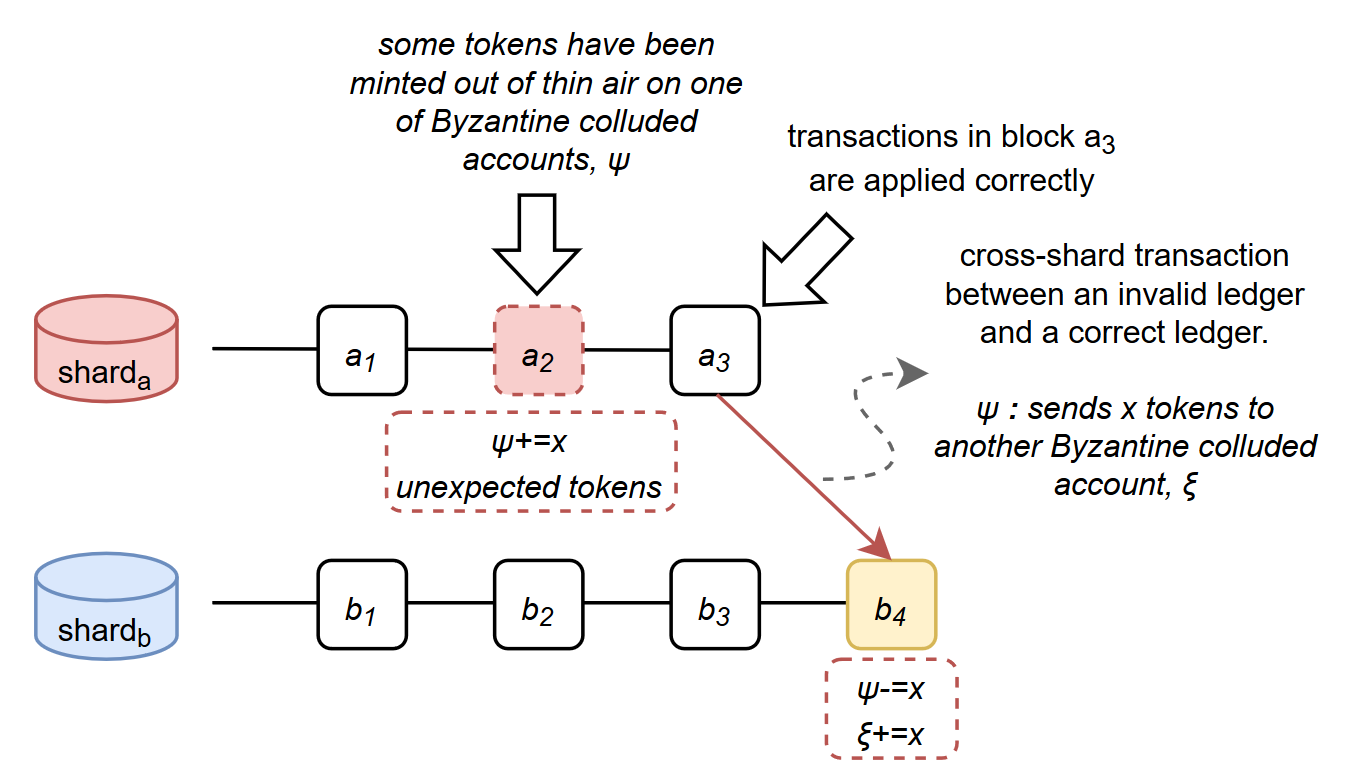}
			\caption{{\footnotesize State transition challenge in sharding.}}
			\label{state_transition-challenge_in-sharding}
		\end{figure} 
		
		\noindent
		- Existing solutions to state transition challenge:
		We first introduce existing solutions for this issue and then analyze their challenges.\vspace{0.1cm}
		
		\begin{itemize}
			\item Processing Preceding Blocks:\\
			\vspace{-3mm} \\
			One of the simplest strategies to defeat the invalid state transition challenge illustrated in Figure \ref{state_transition-challenge_in-sharding} is that processing nodes of $\mathrm{shard_{b}}$ also process the block from which the cross-shard transaction is initiated.
			This approach would not even work in the example depicted in Figure \ref{state_transition-challenge_in-sharding}, because block $a_{3}$ appears to be completely valid.
			As an alternative approach, processing nodes of $\mathrm{shard_{b}}$ should also process some large number of blocks preceding the block from which the cross-shard transaction is initiated. Even this alternative can not be efficient, as for any number of blocks that are validated by $\mathrm{shard_{b}}$, the Byzantine nodes in $\mathrm{shard_{a}}$ can generate one more valid block on top of the invalid block that they created. \\
			
			\item Graph-Based Solution: \\
			\vspace{-3mm} \\
			Another approach to solve the state transition issue in cross-shard transactions is to arrange the shards in an undirected graph so that each shard is connected to several other shards and only cross-shard transactions between neighboring shards are permitted.
			This idea is used in \cite{Chainweb,Zamfir}.
			Cross-shard transactions between non-neighboring shards is routed through multiple shards.
			In this approach, each processing node within a shard is responsible for processing transactions within its own shard as well as transactions from neighboring shards.
			Figure \ref{graph_based_solution_1} depicts such a strategy so that $\mathrm{shard_{b}}$ not only processes its own transactions, but also the transactions of all its neighbors, including $\mathrm{shard_{a}}$.
			Hence, the group of colluding Byzantine processing nodes is not able to finalize the cross-shard transaction depicted in Figure \ref{state_transition-challenge_in-sharding}, because $\mathrm{shard_{b}}$ processes the entire history of $\mathrm{shard_{a}}$ as well---as its neighbor---leading to the identification of invalid block $a_{2}$.
			With the graph-based approach, while corrupting one shard no longer leads to an effective attack, corrupting multiple shards can still be problematic.
			In Figure \ref{graph_based_solution_2}, a colluding Byzantine processing node that has managed to corrupt both $\mathrm{shard_{a}}$ and $\mathrm{shard_{b}}$ can successfully execute a cross-shard transaction to $\mathrm{shard_{c}}$ with funds originating from invalid block $a_{2}$.
			$\mathrm{Shard_{c}}$ processes the entire ledger of $\mathrm{shard_{b}}$, but not that of $\mathrm{shard_{a}}$, and thus it is not able to detect invalid block $a_{2}$.\\ 						
			
			\begin{figure}
				\centering
				\includegraphics[width=1\textwidth]{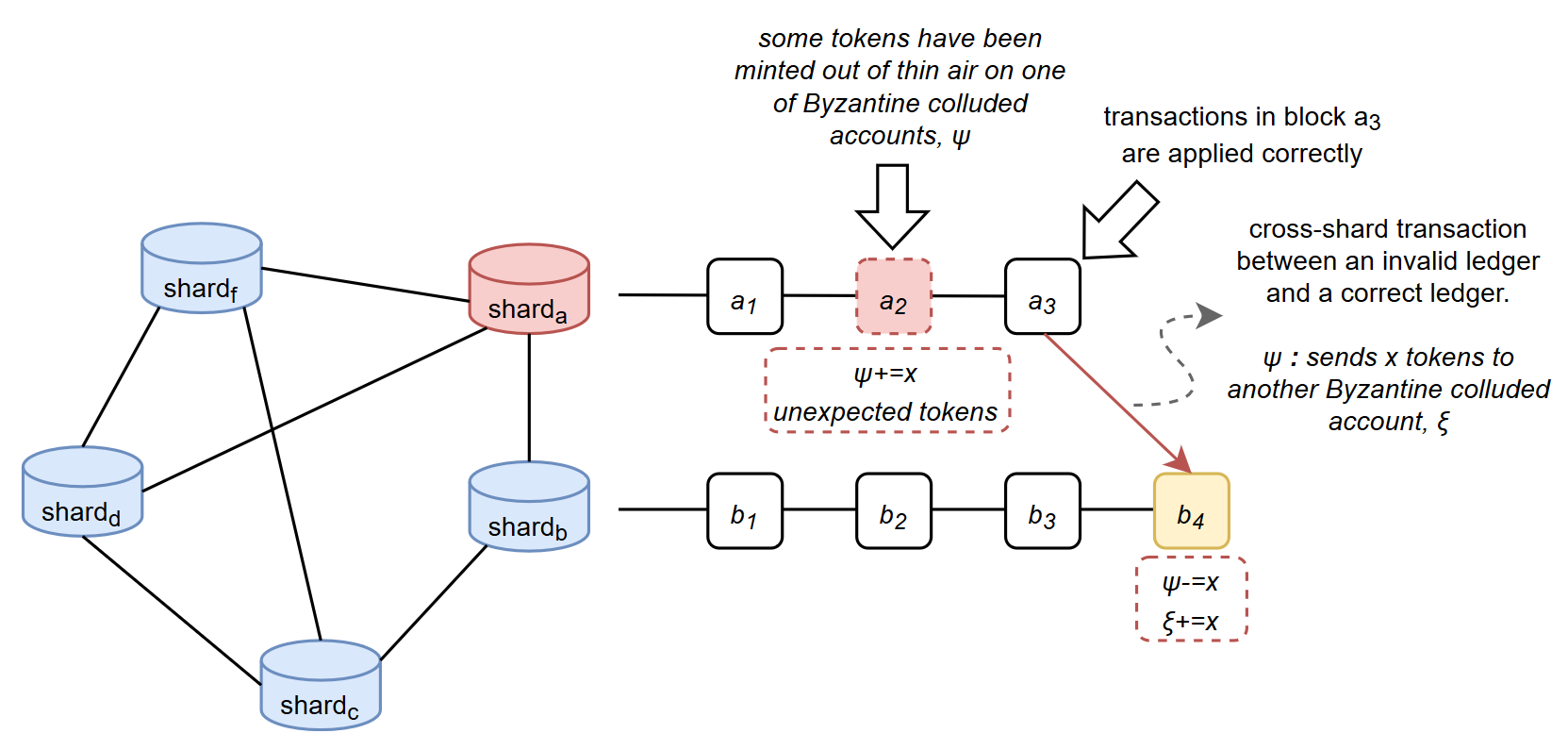}
				\caption{{\footnotesize A graph-based solution can resolve the state transaction problem in some cases.}}
				\label{graph_based_solution_1}
			\end{figure} 
			
			\begin{figure}
				\centering
				\includegraphics[width=1\textwidth]{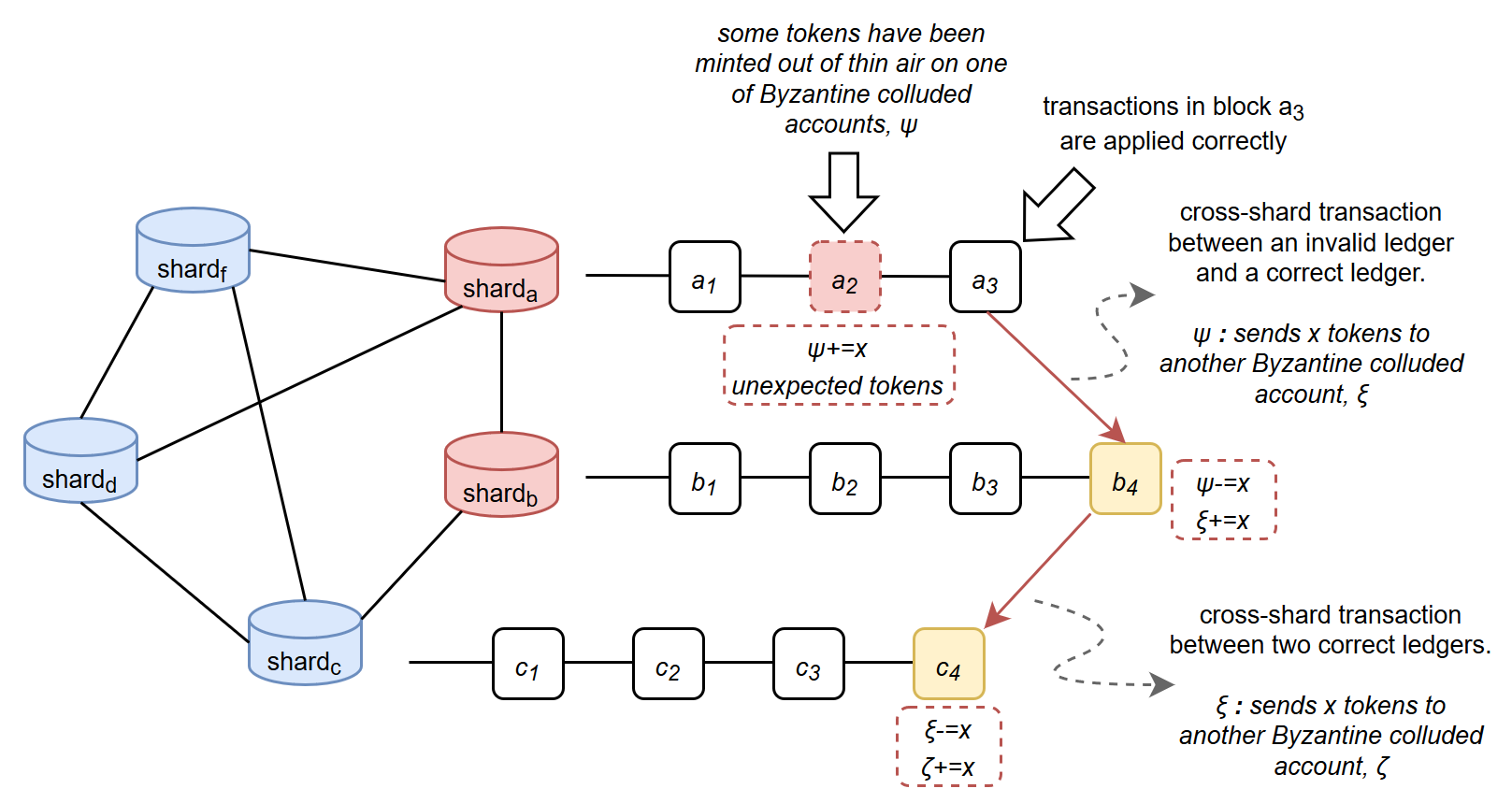}
				\caption{{\footnotesize The graph-based solution is not always able to resolve the state transaction problem.}}
				\label{graph_based_solution_2}
			\end{figure} 
			
			\noindent
			\item Fisherman: \\
			\vspace{-3mm} \\
			The idea behind the Fisherman approach is that whenever a cross-shard transaction occurs, if an honest processing node is present in the shard where the invalid block has been created, within a certain period of time called the challenge period, it can prove that the block is invalid. In fact, a Fisherman node is a member of a shard controlled by a group of colluding Byzantine nodes, and if an invalid block is created by the Byzantine group, it is reported by the Fisherman to the token receiving shard with which a cross-shard transaction is made. As an advantage, this approach works as long as there is at least one honest processing node in the part where the invalid block exists. In order to optimize the communication overhead for the receiving nodes, various constructions are used that enable succinct proof of the invalidity of malicious blocks.
			While this idea represents the prevailing approach in today's proposed protocols, it possesses two notable weaknesses. Firstly, the challenge period must be sufficiently lengthy to allow the honest processor to prepare and thoroughly validate whether a block is invalid. Secondly, considering such a period can substantially decrease the speed of cross-shard transaction processing. Moreover, this approach and the challenge periods required to process invalid blocks create the potential for a new type of attack wherein colluding Byzantine nodes spam the network with invalid challenges.
			A solution for this type of attack is to block some tokens from challenge senders as a deposit or collateral and only return them if the challenge is correct. This solution, however, may not be efficient enough because it may still be profitable for attackers to spam the system and burn its deposits with invalid challenges; for example, when tokens that have been minted out of thin air in the invalid block $a_{2}$ are more than burned collateral tokens. Or, in another scenario, in the case of a griefing attack: an attack that does not necessarily benefit the attacker, but the main motive is to make it harder for the victim to use the system, i.e. to cause him grief, as the name implies.
						
		\end{itemize}
		
	\end{itemize}
	
	\section{Overview of Sharding in Distributed Systems}
	\label{related_works_sharding}
	\noindent
	In this section, we review various replication systems, encompassing both classic distributed databases and Distributed Ledger Technologies (DLTs), that employ the sharding technique.
	Initially, we explore DLT protocols grounded in the sharding concept.
	Two notable sharding protocols are selected for in-depth analysis, representing two general types of ecosystems in sharding techniques: Ethereum 2.0\footnote{Recently, Ethereum 2.0 underwent a change in the architecture of its sharding approach \cite{Ethereum_2_0_New_Sharding_2024}.} \cite{Ethereum_number_of_shards}, one of the most widely used blockchain-based platforms, featuring a homogeneous multi-chain sharding system, and Polkadot \cite{Polkadot}, a heterogeneous multi-chain sharding protocol. Unlike two homogeneous blockchains, two heterogeneous blockchains lack identical architectures and exhibit distinct characteristics.
	After examining these two sharded DLTs, we proceed to review additional DLT sharding protocols in Section \ref{Other_Sharded}, where each aims to address specific challenges in sharding.
	Subsequently, in Section \ref{Classic_Sharding_Databases}, we delve into classic distributed databases. Many prominent distributed databases leverage a combination of sharding and data replication to achieve high availability, fault tolerance, and scalability. We introduce several widely utilized distributed databases that incorporate both sharding and data replication. 
	
	\subsection{Ethereum 2.0: Homogeneous Multi-Chain}
	\label{eth}
	\noindent
	Sharded Ethereum (also known as Ethereum 2.0) \cite{Ethereum_number_of_shards} is an upgraded version of Ethereum that aims at transforming Ethereum into a sharded network.
	Although there are efficient Ethereum clients like Parity \cite{parity} that can process around 3,000 transactions per second, provided the underlying hardware infrastructure is efficient enough, nevertheless, non-sharded Ethereum network is not enabled to process more than about $\approx$ 30 transactions per second \cite{Zilliqa}.
	Hence, Ethereum Foundation decided to upgrade the protocol to a sharding-based system including a multi-phase upgrade to improve Ethereum scalability and capacity.
	In the first phase of the implementation, they set up 64 shards to test the Beacon chain's finality \cite{Polkadot_Wiki,Ethereum_number_of_shards}.
	In the following, we describe the main components as well as major challenges of Ethereum 2.0.\vspace{0.1cm}

\begin{itemize}
	\item Beacon Chain: A Shared Ledger Among Shards:\\
	\vspace{-3mm}\\
	The Beacon chain is the main component of the Ethereum 2.0 network \cite{Ethereum_number_of_shards}.
	Shard chains, on the other hand, are ledgers where transactions are executed \cite{Ethereum_number_of_shards}.
	Each shard chain has an independent state, so it is only responsible for processing transactions related to that state.
	The Beacon chain performs important tasks such as tracking information about validators\footnote{Another term for the nodes in the network which process transactions.}, their stakes, attestations, and votes.
	Additionally, it has the authority to slash validators if they are found to be dishonest, imposing a penalty where dishonest validators lose a portion of their stakes.
	The Beacon chain, as well as its underlying protocol, is responsible for administering consensus between validators on the state of the system.
	Due to the coordination role and the amount of assets managed by the Beacon chain, this ledger is a mission-critical component of the Ethereum 2.0 ecosystem \cite{beacon_chain_verification} and as explained in Section \ref{shared_state_problems}, any bug in this shared ledger could compromise the whole network.\\
	
	\item PoS and Block Generation:\\
	\vspace{-3mm}\\
	In Ethereum 2.0, PoW is replaced by Proof-of-Stake (PoS), an alternative to PoW aimed at reducing computational costs. Proof-of-Stake was first introduced in the Peercoin protocol. \cite{PoS}.
	Ethereum 2.0 finalizes batches of blocks based on time periods called epochs \cite{eth_block_time}.
	An epoch is defined as some constant number of slots. In Ethereum 2.0, time is measured in slots and it is defined as some constant number of seconds.
	In a tentative version, an epoch is considered 64 slots and each slot 12 seconds \cite{Gasper}.
	It is planned to finalize 32 blocks in each batch during one epoch.
	With an estimated block time of 12 seconds, finality is expected to take between 6 to 12 minutes \cite{eth_block_time}.
	To generate blocks, Ethereum 2.0 uses the RandDAO, which is a slot-based protocol that randomly selects validators for a slot and enforces a fork choice rule for unfinalized blocks \cite{BC4T}.
	Each validator instance requires 32 ETH as stake.
	Sets of validators are randomly selected and form groups called committees that validate shards on the network.
	A large number of validators are needed to ensure the validity \cite{Dynamic}.
	A minimum of 111 validators per shard is required to run the network, and 256 validators per shard are required to finalize all shards within one epoch.
	Hence, with 64 shards planned for the first phase, 16,384 validators are needed \cite{Polkadot_Wiki}. \\
	
	\item Roles and Terminology in Ethereum 2.0:\\
	\vspace{-3mm}\\
	In Ethereum 2.0, validators participate in the consensus mechanism and are called virtual miners.
	A proposer is selected pseudo-randomly to propose and build a block for each slot.
	On the other hand, attesters vote on the proposed blocks of both the Beacon chain and the shard chains.
	The vote of the validators is called attestation.
	In most cases, validators are attesters who vote on blocks and their attestations are recorded on the Beacon chain.
	Proposers receive a reward if their proposed block gets confirmation of a quorum of attesters.
	Each attestation has a weight, which is actually the amount of the stake of the validator who writes the attestation \cite{Gasper}.
	This approach is used for the fork choice rule mechanism that we describe in Section \ref{eth_consensus}.
	Validators monitor each other and are rewarded for reporting validators who generate conflicting attestations or propose multiple blocks.
	Each block is proposed by a random block proposer to be added to the Beacon chain in each slot.
	If, for a given slot, a validator does not see a generated block, or does not receive the block in time, or if the block was generated on a chain that the validator does not recognize as the current chain, it should generate an attestation that the slot is empty by attesting to the block it believes is the head of the chain. That is, a validator should attest to exactly one block per slot, either attesting to the actual block generated by a proposer or making an attestation showing that the slot is empty. \\
	
	\item Consensus in Ethereum 2.0:\\
	\vspace{-3mm}\\
	\label{eth_consensus}  
	Gasper, which is a combination of the Casper-FFG and LMD-GHOST, is designed for a full proof-of-stake based blockchain system, where a validator’s voting power is proportional to their stake (or crypto-currency) in the system, such that instead of using computational power to propose blocks, proposing blocks is essentially free \cite{Gasper}.
	Unlike Gasper, Casper-FFG is a hybrid PoW/PoS system.
	It is also based on the consensus theory of Byzantine fault-tolerance \cite{Casper_FFG}.
	In fact, Casper-FFG implements a PoS mechanism as an overlay on top of a PoW ledger to achieve more energy-efficient finality by creating a hybrid consensus model. 
	Casper-FFG is designed to be compatible with a wide range of blockchain protocols with tree-like structure
	and is a \enquote{finality gadget}, meaning that is not a fully-specified protocol and is designed to be a \enquote{gadget} that works on top of a provided blockchain protocol, agnostic to whether the provided chain is proof-of-work or proof-of-stake \cite{Gasper}.
	Casper-FFG is an algorithm that marks certain blocks in a blockchain as finalized so that participants with partial information can still be fully confident that the blocks are part of the canonical chain of blocks \cite{Gasper}.
	Both Gasper and Casper-FFG define the concepts of \enquote{justification} and \enquote{finalization} which are analogous to phase-based concepts in the PBFT literature such as \enquote{prepare} and \enquote{commit} \cite{Gasper}.
	Although in Gasper the \enquote{pairs}\footnote{A pair consists of a block and an epoch.} are justified and finalized rather than the \enquote{checkpoint blocks} in Casper-FFG.
	In order for a block to be \enquote{justified}, two-thirds of all staked ETH must have voted in favor of including that block in the canonical chain.
	If another block is \enquote{justified} above a \enquote{justified} block, that block is upgraded to \enquote{finalized} \cite{Gasper_developers}. 
	Also, in Ethereum 2.0, validators are considered as the same replicas in PBFT \cite{Gasper}. \\
	
	\item Fork Choice Rule in Ethereum 2.0:\\
	\vspace{-3mm}\\
	As transactions throughput accelerates, the probability of the blockchain being forked also increases.
	This may include short-term forks and the possibility of various kinds of censorship.
	The fork choice rule in Ethereum 2.0 is called LMD-GHOST, which stands for Latest Message Driven Greediest Heaviest Observed Subtree.
	In Bitcoin's proof-of-work, the longest chain rule serves as a fork-choice rule that designates the leaf block farthest from the genesis block as the \enquote{heaviest chain} or the \enquote{most difficult chain}.
	However, in Ethereum's proof-of-stake, each attestation carries a weight corresponding to the stake of the validator issuing the attestation.
	This weight serves as a vote, and the fork with the heaviest weight is assumed to be the head of the canonical chain.
	GHOST is a greedy algorithm that grows the blockchain in sub-branches with the \enquote{most activity} \cite{Gasper}.
	LMD-GHOST, a fork-choice rule in Ethereum 2.0, involves validators (participants) attesting to blocks to signal support, similar to voting \cite{Gasper}.
	In LMD-GHOST, the process always converges to a leaf block, defining the canonical chain \cite{Gasper}.
	To define LMD-GHOST, it is necessary to first define a concept called weight.
	Buterin et al. \cite{Gasper} define weight as follows.
	If assume $S$ to be the set of latest attestations, such that one per validator, the weight of block $b$ is defined as the sum of the stake of validators whose last attestation is either to $\beta$ or $\beta$'s  descendants.
	The idea of LMD-GHOST is that at any fork, the subtree of a fork with the heaviest weight is assumed to be the right one, so that it always ends up at a leaf block that defines a canonical chain.
	Figure \ref{LMD_GHOST} illustrates an example of the fork choice rule based on LMD-GHOST.
	Each fork choice in Ethereum 2.0 by a validator $v$ is made in a view at a given time $t$, denoted by $view(v, t)$, as the set of all accepted messages that $v$ has seen so far.
	In a nutshell, based on the LMD-GHOST fork choice rule, anywhere there is a fork, the heaviest subtree is chosen. In this way, in Figure \ref{LMD_GHOST}, the subtree starting with block $\beta_{2}$ is selected because its weight is 9 and greater than the weight of block $\beta_{3}$, which has a weight of 3.
	Then, similarly, among the three children of block $\beta_{2}$, the block with the highest weight, i.e. block $\beta_{4}$, is selected.
	Thus, using the view illustrated in Figure \ref{LMD_GHOST}, a validator will recognize the blue chain as a canonical chain.
	This figure is modeled after the original figure in the article \enquote{Combining GHOST and Casper} \cite{Gasper} where the LMD-GHOST protocol has been detailed.
	The main idea of GHOST is to choose a side with more overall support for validators in each fork instead of choosing a subtree that is longer, so that in addition to the number of attestations, the weight of each attestation, which is based on the stake of its validator, is also considered.
	This idea is heavily inspired by Sompolinsky et al \cite{GHOST} in the original GHOST paper,
	but LMD-GHOST adapts the design from its original PoW context to new PoS context \cite{Ethereum_2_Phase_0_Github}.
	
	\begin{figure}
		\centering
		\includegraphics[width=0.58\textwidth]{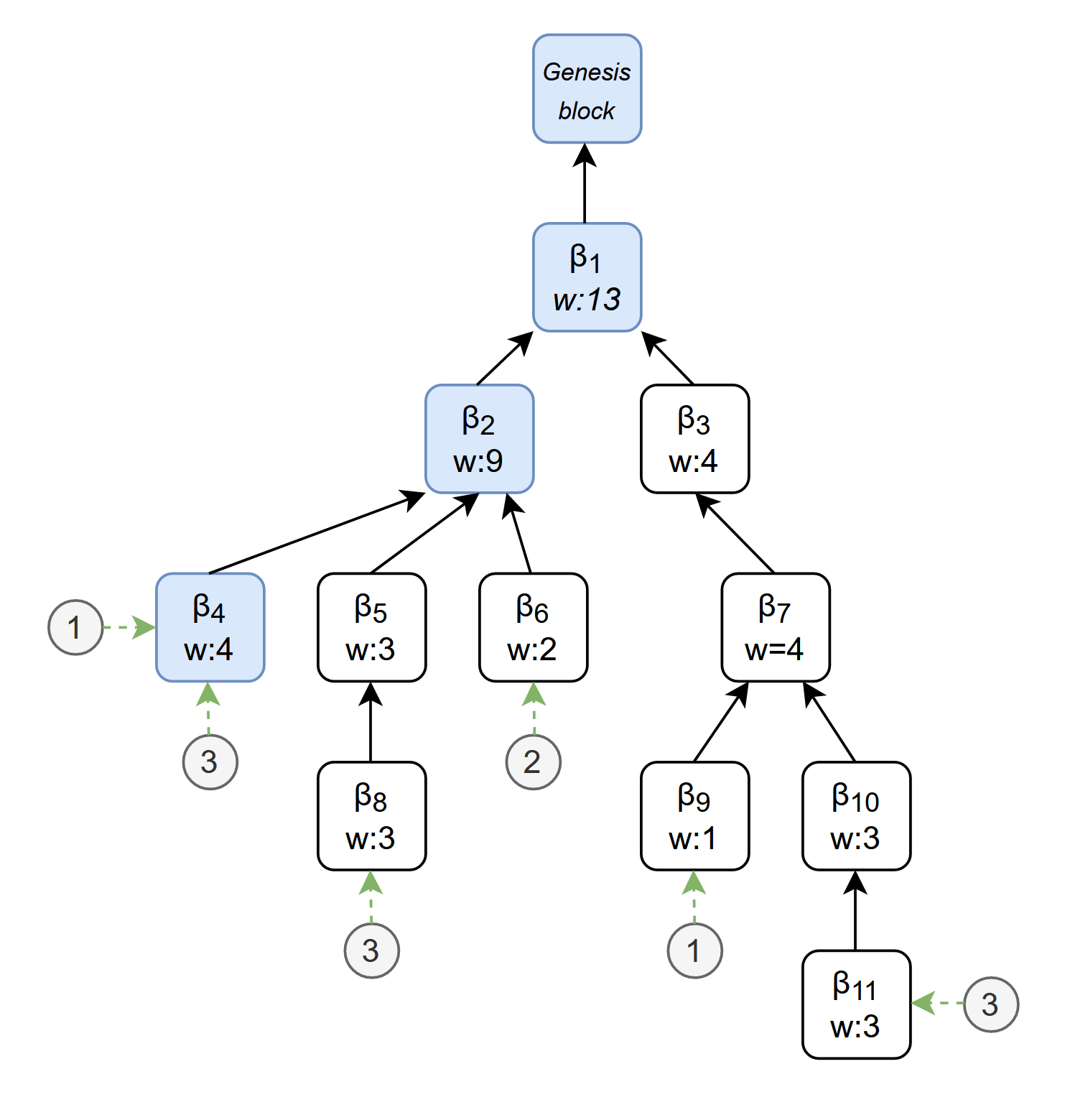}
		\caption{{\footnotesize An example of how to select a subtree using the LMD-GHOST fork choice rule in a view by a validator.
		}}
		\label{LMD_GHOST}
	\end{figure}
	
\end{itemize}
	
	\subsection{Polkadot: Heterogeneous Multi-Chain}
	\label{Polkadot}
	\noindent
	Polkadot was first introduced by Gavin Wood in 2016 as a heterogeneous multi-chain protocol aiming to provide a scalable and interoperable framework for multiple chains with pooled security that is achieved by the collection of components \cite{Polkadot}.
	Polkadot has its own native crypto-token called DOT.
	Polkadot uses a central chain called the \textit{Relay} chain that communicates with several heterogeneous and independent chains called \textit{parachains}.
	The Relay chain is responsible for providing shared security to all parachains as well as enabling trustless cross-chain transactions between parachains.
	The issues Polkadot intends to address are interoperability, scalability, and weaker security resulting from splitting the security power \cite{Polkadot_overview}.
	The Polkadot Relay chain consists of nodes and roles.
	Nodes are network-level entities that physically run Polkadot software, and roles are protocol-level entities that administer specific purposes.
	At the network level, Relay chain nodes can participate as either light clients or full nodes.
	Unlike light clients, which retrieve certain user-relevant data from the network and are not required to be always available because they do not perform a service for others, full nodes retrieve all types of data. They store this data for a long time, disseminating it to others, and therefore should be highly available \cite{Polkadot_overview}.
	In addition to data distribution, Relay chain nodes perform specific roles at the protocol level as follows:\vspace{0.1cm}
	
	\begin{itemize}
		\item Validators: as the Relay chain full nodes, do the bulk of the security work.\vspace{0.1cm}
		\item Nominators: are shareholders who elect the validator candidates. This can be done through a light client without need of any awareness of parachains.\\
	\end{itemize}
	
	\vspace{-0.1cm}
	\noindent On the other side, parachains can determine their internal network structure but are expected to interact with Polkadot through the following roles:\vspace{0.1cm}
	\begin{itemize}
		\item Collectors: collect parachain data and send it to the Relay chain. Collectors are selected as defined by parachain and must be its full nodes. Validators interact with parachain collators, but do not need to participate in parachain as a full node.\vspace{0.1cm}
		\item Fishermen: as we explained previously, a Fisherman can assist with data validity challenges. As a full node of the parachain, they perform additional security checks to ensure the correct functioning of the parachain on behalf of the Relay chain. In return, the Relay chain provides rewards to incentivize the Fishermen.\\
	\end{itemize}
	
	\vspace{0.1cm}
	\noindent In addition to the components listed above, the bridges are intended for compatibility and interaction of the Polkadot ecosystem with other external blockchain systems such as Bitcoin, Ethereum or Tezos \cite{Polkadot_Wiki}. \\
	
\begin{figure}
		\centering
		\includegraphics[width=0.75\textwidth]{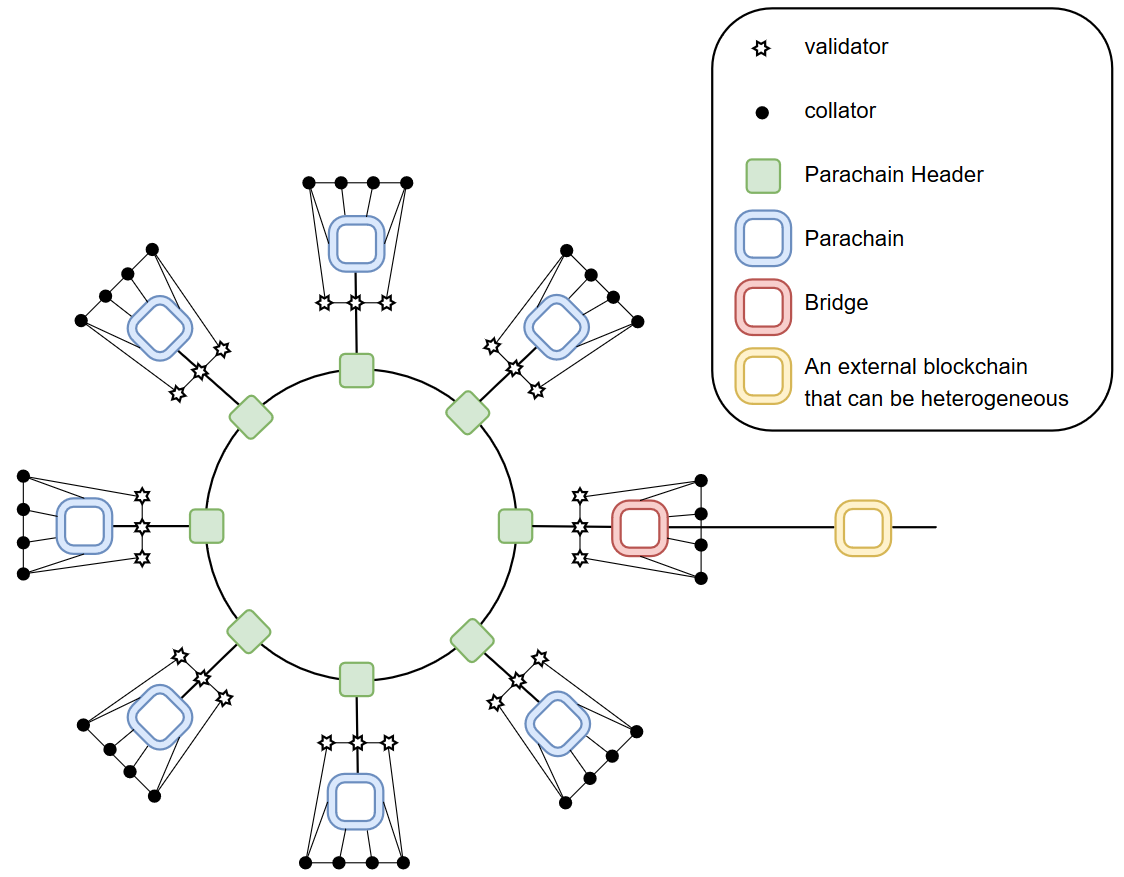}
		\caption{{\footnotesize A high-level view of the Polkadot architecture.}}
		\label{polkadot_figure}
	\end{figure}

	\noindent After describing the nodes and roles, we explain more details of how the Polkadot Relay chain protocol works as follows.
	Collaborators watch the progress of the block-producing and consensus protocols. They sign the data building on top of the latest chain block and send it to the validators assigned to their parachain in order to include it in the Relay chain.
	The parachain validators decide which of the parachain block to support in order to present its relevant data as a parachain next candidate for being added to the next Relay chain block.
	A block-producing validator makes a set of candidates from all parachains and puts it into a Relay chain block.
	Validators send their votes on a block and finalize it. All votes are included in the Relay chain blocks.
	Figure \ref{polkadot_figure} depicts a high-level view of the Polkadot architecture in an example that includes 8 parachains, 24 validators, 4 collators per parachain, along with a bridge connecting the entire system to another external blockchain network so that two blockchain networks can be heterogeneous.
	This figure is modeled after the figure in \cite{Polkadot_overview}, where the Polkadot protocol is thoroughly reviewed.\\
	
	
	\noindent As a consensus mechanism, Polkadot uses Nominated PoS that is a modified version of proof-of-stake.
	Since NPoS has a deterministic finality, a set of registered validators of bounded size is required.
	DOT holders can participate in the NPoS consensus as nominators.
	To register as a nominator, at least 100 DOT is required.
	The validators' candidacies are visible to all nominators. Each nominator publishes a list of up to 16 candidates it supports. The network then automatically distributes the stake among the validators based on the nominations.
	And finally, a certain number of validators that have the most DOT (as stake) are selected and activated. 
	In NPoS, the stake of nominators and validators may be slashed, as a security measure \cite{Polkadot_Wiki}.
	
	\subsection{Other Sharded Blockchains}
	\label{Other_Sharded}
	\noindent
	In this section, we review some other notable sharding protocols, each of which attempts to ameliorate some of the sharding challenges.\\
	
	\noindent
	Zilliqa \cite{Zilliqa} was proposed as a sharding-based model for permissionless blockchain networks in order to improve the scalability issues.
	A major weakness of Zilliqa is that it shards processing but not storage \cite{Nightshade}, that is, each node holds the entire stored replicated data state to be able to process transactions \cite{Nightshade}.
	It is worth noting that the decision not to shard by state, while simplifies the system design, imposes a huge limit on the scalability of the system \cite{zilliqa_limitations}.
	In fact, only supporting processing sharding prevents machines with limited resources from participating in the network, thus curtailing decentralization \cite{Harmony_protocol}.
	Zilliqa uses PoW as an identity registration process as a Sybil attack \cite{sybil} mitigation mechanism.
	Zilliqa's consensus core relies on PBFT, improving its efficiency using EC-Schnorr multi-signature as developed in \cite{honest_or_bust} and \cite{enhancing_bitcoin_security}.\\
	
	\noindent
	Elastico \cite{Elastico} is proposed as a sharding-based protocol for permissionless blockchain networks and uniformly partitions network into smaller committees, each of which processes a disjoint set of transactions. According to the results of their experiments to measure the scalability of the network based on PBFT consensus, when network size increases from 40 to 80 nodes (2 times), the latency to reach consensus for each transaction is 6 times longer (e.g. from 3 seconds to 18 seconds), and their experiment did not terminate even after running for 1 hour with a network size of 320 nodes.\\
	
	\noindent
	Omniledger \cite{OmniLedger} is another sharding-based solution, whose construction is close to Elastico, but it brings up some challenges in Elastico and then targets to solve them. In Omniledger, the validators\footnote{Another term for the nodes in the network which process transactions.} are selected by use of proof-of-work. In this way, they use a sliding window of latest block miners as the validator set. They also utilize proof-of-stake as an alternative Sybil attack resistant approach for choosing a set of validators, aiming to achieve a more power-efficient consensus.
	They implemented a prototype in Go language on commodity servers (12-core VMs on Deterlab) and their experimental results show that OmniLedger throughput is 6,000 transactions per second with a 10 second consensus latency for 1,800 nodes.\\
	
	\noindent
	SharPer \cite{Sharper} is a permissioned blockchain system in which nodes are clustered and each data shard is replicated on the nodes of a cluster so that each cluster maintains only a view of the ledger. In SharPer, the blockchain ledger is formed as a DAG. \\
	
	\noindent
	Ren et al. \cite{spontaneous_sharding} introduce a permissioned blockchain and call it \enquote{spontaneous sharding} so that the network consists of three parts: (a) individual chains generated by each node to record their own transactions in a first-in-first-out fashion, (b) a main chain for a global shared state that uses PBFT as its consensus algorithm and the blocks consist of abstracts signed by the corresponding nodes. They assume that the abstracts of all genesis blocks are on the main chain. Honest nodes will send abstracts of their newest blocks to the main chain when they observe that their previous abstracts are on-chain, and (c) a validation scheme for validation of the transactions. The transactions on individual chains are arbitrary in the sense that they are neither tamper-proof nor signed. The transactions will be tamper-proof and signed if an abstract of a block that comes after it is contained in the main chain, which they call \enquote{confirmed} transactions. \\
	
	\noindent
	Dang et al. \cite{Towards_sharding} propose a sharding-based system for permissioned blockchains relying on a trusted execution environment, namely Intel SGX to eliminate equivocation in the Byzantine failure model. As the authors claim, without equivocation, existing BFT protocols can achieve higher fault tolerance with the same number of nodes, that is, a committee of $n$ nodes can tolerate up to $(n-1)/2$ non-equivocating Byzantine failures, as opposed to $(n-1)/3$ failures in the original threat model \cite{Hybrids_on_steroids,Attested_append_only_memory,TrInc}. They have run their experiments on a local cluster with 100 nodes consisting of over 1,400 Google Cloud Platform (GPC) nodes distributed across 8 regions.
	On GPC setup, they achieved a throughput of over 3,000 transactions per second.
	
	\subsection{Sharding in Classic Distributed Databases}
	\label{Classic_Sharding_Databases}
	\noindent
	Several prominent distributed databases leverage a combination of sharding and data replication to attain high availability, fault tolerance, and scalability.
	Below, we introduce several widely utilized distributed databases that incorporate both sharding and data replication. \vspace{0.1cm} 
	
	\begin{itemize}
		\item Apache Cassandra:\\
		\vspace{-3mm} \\
		\noindent
		Apache Cassandra is a distributed NoSQL database designed for handling large amounts of data across multiple commodity servers without a single point of failure.
		Cassandra operates on a peer-to-peer architecture where all nodes in the cluster are treated equally.
		Each node in the cluster is responsible for a portion of the data, and there is no central coordinator.
		In Apache Cassandra, data distribution across the cluster is achieved using a consistent hashing algorithm.
		This algorithm is non-cryptographic in nature and is designed to evenly distribute data across the nodes in the cluster.
		Consistent hashing helps in ensuring a balanced distribution of data while allowing for easy addition or removal of nodes in the cluster without significant reorganization of data.
		Each node is assigned a range of the hash function, and this helps in evenly distributing data across the nodes.
		Cassandra ensures fault tolerance through data replication.
		Each piece of data is replicated across multiple nodes (data centers) to ensure high availability and fault tolerance.
		Write operations involve writing data to the node responsible for the partition determined by the hash of the partition key.
		Read operations can be served by any node in the cluster, as data is replicated.
		Cassandra allows users to configure the consistency level for read and write operations.
		Consistency levels determine how many nodes in the cluster need to acknowledge a read or write for it to be considered successful.
		Cassandra provides tunable consistency, allowing users to balance between consistency and availability based on the application's requirements.\\
		
		\noindent
		While Apache Cassandra is a powerful and scalable NoSQL database, it does have some weaknesses that should be considered:\vspace{0.1cm}
		
		\begin{itemize}
			\item Setting up and configuring Cassandra can be complex, especially for those new to distributed databases.\vspace{0.2cm}
			
			\item Fine-tuning parameters and understanding the impact of configuration changes may require expertise.\vspace{0.2cm}
			
			\item Cassandra uses its own query language, CQL (Cassandra Query Language), which lacks some advanced querying features compared to SQL.\vspace{0.2cm}
			
			\item Complex queries involving multiple tables or joins are not as straightforward as in relational databases.\vspace{0.2cm}
			
			\item Cassandra prioritizes high availability and scalability over strong consistency, leading to limited support for ACID transactions.\vspace{0.2cm}
			
			\item It follows the eventual consistency model, which may not be suitable for use cases requiring strict transactional guarantees.\\
		\end{itemize}
		
		\vspace{-0.2cm}
		\noindent
		Additional information about Apache Cassandra can be explored in various sources, including \cite{Cassandra_book_1,Cassandra_book_2,Cassandra_documentation,Cassandra_paper}.\\
		
		\item Amazon DynamoDB:\\
		\vspace{-3mm} \\
		\noindent
		Amazon DynamoDB is a fully managed NoSQL database service designed for high-performance and scalable applications.
		Sharding in DynamoDB refers to the process of partitioning a table's data across multiple physical storage partitions called shards.
		Each shard is an independent data store that can handle a specific amount of read and write capacity.
		DynamoDB uses a table's partition key to distribute data across shards.
		Items with the same partition key are stored together in the same shard.
		DynamoDB replicates data across multiple Availability Zones (AZs) to ensure durability and high availability.
		It also offers backup and restore capabilities for data protection.
		Availability Zones are distinct data center locations within a region that are designed to be isolated from each other.
		Amazon Web Services (AWS) has built its infrastructure to include multiple Availability Zones in each AWS region to provide customers with increased fault tolerance and high availability.
		Sharding in DynamoDB refers to the process of partitioning a table's data across multiple physical partitions, known as shards.
		Each shard is an independent storage unit with its own provisioned throughput capacity.
		DynamoDB partitions data across multiple servers to handle large datasets and high read/write throughput.
		Each partition is known as a partition key range and is associated with a specific hash value.
		The partition key is a crucial concept in DynamoDB sharding.
		It determines the partition (or shard) in which the item is stored.
		Well-chosen partition keys are essential for even distribution of data and optimal query performance.
		Further insights into DynamoDB are available in diverse references, such as \cite{Dynamo,DynamoDB,DynamoDB_book_1,DynamoDB_book_2,DynamoDB_book_3}.\\
		
		\item Google Bigtable:\\
		\vspace{-3mm} \\
		\noindent
		Google Bigtable is a highly scalable and fully managed NoSQL database service designed for large-scale, real-time applications.
		Developed by Google, it provides a distributed storage system that can handle massive amounts of data across multiple servers.
		Bigtable is particularly well-suited for applications requiring high throughput and low-latency access to vast amounts of structured data, such as analytics, IoT, and financial services.
		It employs a sparse, distributed, and persistent multi-dimensional sorted map data structure, making it efficient for dynamic and evolving datasets. Bigtable's automatic sharding and replication capabilities ensure high availability, fault tolerance, and seamless scalability, making it a robust choice for organizations with demanding performance and scalability requirements.
		More details regarding Google Bigtable can be investigated through a variety of sources, including \cite{Bigtable_Documentation,Google_Cloud_Whitepapers}.\\
		
		\item Google Spanner:\\
		\vspace{-3mm} \\
		\noindent
		Google Spanner is a globally distributed, strongly consistent, and horizontally scalable database service developed by Google. It combines the benefits of traditional relational databases with the flexibility of NoSQL databases. Spanner is designed to provide seamless global transactions across multiple data centers, ensuring high availability and low-latency access to data. It uses a unique combination of synchronized clocks and a two-phase commit protocol to achieve external consistency, making it suitable for applications requiring strong data consistency in a distributed environment. Spanner's architecture allows it to automatically shard data and scale horizontally, enabling it to handle large workloads across the globe while maintaining ACID properties.
		Diverse references, including \cite{Google_Spanner_original_paper,Google_Cloud_Spanner_Documentation,Spanner_Becoming_a_SQL_system}, offer further insights into Google Spanner.\\
		
		\item ScyllaDB:\\
		\vspace{-3mm} \\
		\noindent
		ScyllaDB is a highly performant and scalable NoSQL database designed for maximum efficiency in handling large volumes of data and high-throughput workloads.
		ScyllaDB leverages a shared-nothing architecture and is implemented in C++ for enhanced performance.
		A shared-nothing architecture is an architectural design where individual nodes in a distributed system operate independently and do not share any physical components, such as memory or storage, with each other. 
		It excels in real-time big data applications, providing low-latency, fault-tolerance, and linear scalability across distributed environments.
		Its design prioritizes simplicity, robustness, and ease of integration, making it a compelling choice for organizations requiring a reliable and high-performance solution for their data storage and retrieval needs.
		ScyllaDB is designed to be compatible with Apache Cassandra at the application level. In other words, if an application is already using Apache Cassandra as its database, it is possible to replace it with ScyllaDB without making significant changes to the application code.
		There is a wealth of information about ScyllaDB in various sources, including \cite{ScyllaDB_Documentation,ScyllaDB_GitHub_Repository}.\\
		
		\item MongoDB:\\
		\vspace{-3mm} \\
		\noindent
		MongoDB \cite{MongoDB} is a NoSQL database management system that provides a scalable and flexible platform for handling large volumes of unstructured or semi-structured data. It falls under the category of document-oriented databases, where data is stored in a flexible, JSON-like format called BSON (Binary JSON).
		Sharding is a crucial concept in MongoDB that enables horizontal scaling of the database to handle large amounts of data and high read and write throughput. In MongoDB, sharding involves distributing data across multiple machines, called shards, to improve performance and manageability.\\

		\noindent
		Outlined is a detailed explanation of the sharding concept in MongoDB:\vspace{0.1cm}
		
		\begin{itemize}
			\item Shard:
			A shard is a separate MongoDB instance or server that stores a subset of the data. Each shard holds a portion of the entire dataset. Shards can be physical servers or virtual machines.\vspace{0.2cm}
			
			\item Shard Key:
			The shard key is a field or set of fields chosen to distribute data across shards. MongoDB uses the values of the shard key to determine the target shard for each document. The choice of a good shard key is crucial for efficient sharding.
			Choosing an appropriate shard key is critical for the efficiency of sharding. Ideally, the shard key should distribute data evenly across shards, avoiding hotspots and ensuring a balanced workload.\vspace{0.2cm}
			
			\item Shard Cluster:
			The entire distributed database system, consisting of multiple shards, is referred to as a shard cluster. A shard cluster includes a query router (`mongos'), which acts as an interface between the application and the individual shards.\vspace{0.2cm}
			
			\item Query Router (`mongos'):
			The query router, or `mongos', is a routing service that directs queries to the appropriate shard based on the shard key. Applications connect to the query router rather than directly to individual shards, allowing for a unified view of the entire dataset.\vspace{0.2cm}
			
			\item Config Servers:
			MongoDB uses config servers to store metadata about the distribution of data across shards. The config servers maintain information about the ranges of shard key values associated with each shard.\vspace{0.2cm}
			
			\item Chunks:
			The data in a sharded MongoDB database is divided into chunks, which are contiguous ranges of shard key values. Each chunk is stored on a specific shard. As the data size grows or decreases, MongoDB automatically migrates chunks between shards to maintain an even distribution of data.\vspace{0.2cm}
			
			\item Balancing:
			MongoDB's balancer is responsible for redistributing chunks across shards to ensure an even distribution of data. The balancer runs in the background and uses the config servers to determine when and where to move chunks.\vspace{0.2cm}
			
			\item Adding and Removing Shards:
			MongoDB supports dynamic scaling by allowing the addition or removal of shards without downtime. When adding a new shard, the balancer redistributes chunks to balance the data load. Removing a shard involves redistributing its data to the remaining shards.\\
		\end{itemize} 
		
		\item Apache HBase:\\
		\vspace{-2mm} \\
		\noindent
		HBase \cite{HBase} is a distributed, scalable, and consistent NoSQL database that is designed to handle large amounts of sparse data. It is part of the Apache Hadoop project.
		HBase is suitable for storing and managing vast amounts of data across clusters of commodity hardware.
		It provides real-time read and write access to your data and is particularly well-suited for applications where quick and random access to large amounts of data is essential.
		HBase uses a column-family-based data model, similar to Google Bigtable \cite{Bigtable_Documentation}, where data is organized into tables with rows identified by a unique key. Each table can have multiple column families, and each column family can have multiple columns. This schema flexibility allows for efficient storage and retrieval of data.
		A column-family-based data model is a type of NoSQL data model used by certain distributed databases, including HBase. This model is inspired by Google's Bigtable and is designed to provide flexibility in handling large amounts of data with a dynamic schema.
		In a column-family-based model, data is organized into tables. Unlike traditional relational databases with fixed columns, this model groups columns into families, each capable of accommodating multiple columns. Within a family, columns are stored together on disk, promoting cohesive access. This approach allows for the systematic organization of related data and establishes a specific structural framework within a table.
		In HBase, sharding is the process of partitioning and distributing the data across the cluster to achieve scalability and parallelism. The primary goal of sharding is to ensure that the workload is evenly distributed among the region servers in the HBase cluster, preventing hotspots and optimizing performance.\\
		
		\noindent
		Presented are key concepts related to the sharding concept in HBase:\vspace{0.1cm}
		
		\begin{itemize}
			\item Regions:
			A region is a contiguous range of rows in an HBase table. Each region is assigned to a specific region server in the cluster. As the table grows, HBase automatically splits regions to maintain a balanced distribution of data.\vspace{0.2cm}
			
			\item Row Key and Region Assignment:
			The row key plays a crucial role in sharding. It is used to determine the placement of data within regions. HBase uses hashing on the row key to distribute rows across regions. The hash value of the row key determines the region to which the data belongs.\vspace{0.2cm}
			
			\item Automatic Region Splitting:
			As data in a region grows beyond a certain threshold (configured by the HBase administrator), HBase automatically triggers a process called region splitting. Region splitting divides a large region into two smaller, roughly equal-sized regions. Each of these new regions is then assigned to different region servers.\vspace{0.2cm}
			
			\item Zookeeper Coordination:
			Apache ZooKeeper\footnote{ZooKeeper \cite{Zookeeper} is a distributed coordination service that provides primitives for building consensus mechanisms in distributed systems. The consensus protocol used by ZooKeeper is known as the ZAB (ZooKeeper Atomic Broadcast) protocol \cite{Zab}.} is used by HBase for coordination and management of distributed processes.
			ZooKeeper plays a role in maintaining metadata about the state of regions and region servers in the HBase cluster.
			It helps in coordinating tasks such as region assignments and tracking the health of region servers.\vspace{0.2cm}
			
			\item HBase Master:
			The HBase master node is responsible for overall coordination and management of the HBase cluster.
			It assigns regions to region servers, monitors their status, and takes corrective actions, such as initiating region splits or migrations, to maintain cluster health.\\
		\end{itemize}
		
		\item Riak:\\
		\vspace{-2mm} \\
		\noindent
		Riak \cite{Riak_handbook} is a distributed NoSQL database designed to provide high availability, fault tolerance, and scalability for handling large amounts of data across multiple nodes. It was developed by Basho Technologies and is based on the principles of Amazon's Dynamo \cite{Dynamo}, a highly available key-value storage system.
		In Riak, the sharding concept is implemented through the use of consistent hashing. 
		Here's a breakdown of how the sharding process works in Riak:\vspace{0.1cm}
		
		\begin{itemize}
			\item Key Space Partitioning:
			Riak uses a consistent hashing algorithm to map keys to partitions (shards). This algorithm ensures that keys are evenly distributed across the available partitions, preventing hotspots and balancing the load.\vspace{0.2cm}
			
			\item Virtual Nodes:
			To enhance flexibility and manageability, each physical node in the Riak cluster is divided into multiple virtual nodes. Each virtual node is responsible for a subset of the overall key space. This division allows for more granular control over data distribution and enables dynamic scaling.\vspace{0.2cm}
			
			\item Repartitioning:
			When the cluster size changes (nodes are added or removed), Riak can dynamically repartition the key space to ensure an even distribution of data. This automatic repartitioning helps maintain load balance and optimal performance.\vspace{0.2cm}
			
			\item Quorum-based Operations:
			Riak employs a quorum-based system for read and write operations. Quorums define the number of nodes that must participate in an operation for it to be considered successful. This approach enhances fault tolerance and consistency in the presence of network partitions or node failures.\\\\
		\end{itemize}
		
		\item Couchbase:\\
		\vspace{-2mm} \\
		\noindent
		Couchbase \cite{Couchbase_Documentation,Couchbase_analytics} is a NoSQL database that is designed to handle large amounts of unstructured or semi-structured data across multiple nodes in a distributed architecture.
		Couchbase operates in a cluster, which is a group of nodes that work together to store and manage data.
		Each server in the Couchbase cluster is called a node. Nodes can be added or removed dynamically to scale the cluster.
		Couchbase is primarily a key-value store, where data is stored in the form of key-value pairs. The keys are unique identifiers for the data, and values can be JSON documents, binary data, or other formats.
		Data is organized into buckets, which are logical containers for documents. Each bucket can have its own configuration and can be considered as a separate namespace for documents.
		Couchbase uses N1QL, a SQL-like query language, to query JSON documents. N1QL supports both ad-hoc and prepared queries.
		Couchbase implements sharding through its consistent hashing mechanism and a feature known as vBuckets (virtual buckets).
		Here's a detailed explanation of how sharding is designed and implemented in Couchbase:\vspace{0.1cm}
		
		\begin{itemize}
			\item Key-Based Sharding:
			Couchbase uses a key-based sharding approach, where data is distributed across nodes based on the document key. This ensures that related data is stored on the same node, reducing the need for cross-node communication during query operations.\vspace{0.2cm}
			
			\item Consistent Hashing:
			Couchbase employs consistent hashing to distribute keys across nodes in a deterministic manner. Consistent hashing ensures that when the number of nodes in the cluster changes, only a minimal amount of data needs to be relocated, minimizing the impact on the system.\vspace{0.2cm}
			
			\item vBuckets (Virtual Buckets):
			Couchbase divides the data into smaller units called vBuckets. Each vBucket is assigned to a specific node in the cluster. This provides a finer level of granularity for data distribution and ensures that the load is evenly balanced among nodes.\vspace{0.2cm}
			
			\item Replication:
			To enhance data availability and fault tolerance, Couchbase uses replication. Each vBucket has one or more replicas, and these replicas are stored on different nodes. If a node fails, the system can continue to serve data from the replicas on other nodes.\vspace{0.2cm}
			
			\item Automatic Rebalancing:
			Couchbase supports automatic rebalancing, allowing the addition or removal of nodes from the cluster without manual intervention. During rebalancing, the system redistributes vBuckets to maintain a balanced load across nodes.
		\end{itemize}
		
	\end{itemize}
	
	\section{Conclusion}
	\noindent
	In this article, we described the most important challenges in the sharding of distributed replication systems.
	We explained why most current sharded Distributed Ledger Technology (DLT) protocols use a random assignment approach for allocating and distributing nodes between shards due to security reasons.
	We detailed how a transaction is processed in sharded DLTs based on current sharding protocols.
	We also described how a shared ledger among shards imposes additional scalability limitations and security issues on the network.
	Additionally, we explained why cross-shard or inter-shard transactions are undesirable and more costly, due to the problems they cause, including atomicity failure and state transition challenges, along with a review of proposed solutions.
	Furthermore, we reviewed the most important and well-known replication systems, encompassing both classic distributed databases and Distributed Ledger Technologies, which employ the sharding technique.
	
	\ifCLASSOPTIONcaptionsoff
	\newpage
	\fi


\begin{thebibliography}{1}
		
		\bibitem{PBFT}
		Castro, Miguel.  "Practical Byzantine fault tolerance." Ph.D. Dissertation. Massachusetts Institute of Technology. Laboratory for Computer Science. Cambridge, Massachusetts, USA. January 31, 2001. Available online at: \url{https://pmg.csail.mit.edu/~castro/thesis.pdf}
		
		\bibitem{Paxos}
		Leslie Lamport. The part-time parliament. ACM Transactions on Computer Systems, 16(2):133–169, May 1998.
		
		\bibitem{Raft}
		Ongaro, Diego, and John Ousterhout. "In search of an understandable consensus algorithm." 2014 USENIX Annual Technical Conference (Usenix ATC 14). 2014.
		
		\bibitem{Hotstuff}
		Yin, Maofan, et al. "Hotstuff: Bft consensus with linearity and responsiveness." Proceedings of the 2019 ACM Symposium on Principles of Distributed Computing. 2019.
		
		\bibitem{blockchain_consensus_protocols}
		Alqahtani, Salem, and Murat Demirbas. "Bottlenecks in blockchain consensus protocols." 2021 IEEE International Conference on Omni-Layer Intelligent Systems (COINS). IEEE, 2021.
		
		\bibitem{Bitcoin}
		Nakamoto, Satoshi. Bitcoin: A peer-to-peer electronic cash system. Manubot, 2019.
		
		\bibitem{Zilliqa}
		Team, Zilliqa. "The zilliqa technical whitepaper." Retrieved September 16 (2017): 2019.
		
		\bibitem{pow_is_not_consensus_1}
		Ethereum community. Consensus mechanism, Sybil resistance \& chain selection.
		\enquote{PoW and PoS alone are not consensus protocols, but they are often referred to as such for simplicity.}
		Available onlaine at \url{https://ethereum.org/en/developers/docs/consensus-mechanisms/#sybil-chain}.
		Also available in the Internet Archive at: \url{https://web.archive.org/web/20230727221549/https://ethereum.org/en/developers/docs/consensus-mechanisms/}
		
		\bibitem{pow_is_not_consensus_2}
		Gün Sirer, Emin. 
		\enquote{there is a terribly wrong framework emerging around consensus protocols. People think that PoW and PoS are consensus protocols, and that they are the only two consensus protocols out there. This is false.} 13 jun 2018. Tweet.
		Also available in the Internet Archive at: \url{https://web.archive.org/web/20230728023656/https://twitter.com/el33th4xor/status/1006931658338177024}
		
		\bibitem{pow_is_not_consensus_3}
		Zhelezov, Dmitrii. \enquote{PoW, PoS and DAGs are NOT consensus protocols.} (2018).
		Medium. Available online at: \url{https://medium.com/coinmonks/a-primer-on-blockchain-design-89605b287a5a}
		
		\bibitem{pow_is_not_consensus_4}
		Beyer, S. \enquote{Proof-of-Work Is Not a Consensus Protocol: Understanding the Basics of Blockchain Consensus.} Medium. Available online at: \url{https://medium.com/cryptronics/proof-of-work-is-not-a-consensus-protocol-understanding-the-basics-of-blockchain-consensus-30aac7e845c8} (accessed April 1, 2019) (2019).
		
		\bibitem{Hyperledger}
		Dhillon, Vikram, David Metcalf, and Max Hooper. "The hyperledger project." Blockchain enabled applications. Apress, Berkeley, CA, 2017. 139-149.
		
		\bibitem{Database_system_concepts}
		Silberschatz, Abraham, Henry F. Korth, and Shashank Sudarshan. Database system concepts. McGraw-Hill, 2011.
		
		\bibitem{proof-of-x}
		The Bitcoin Wiki, Category: Proof-of-x. Available online at: \url{https://en.bitcoin.it/wiki/Category:Proof-of-x}
		
		\bibitem{eth_tps}
		Bez, Mirko, Giacomo Fornari, and Tullio Vardanega. "The scalability challenge of ethereum: An initial quantitative analysis." 2019 IEEE International Conference on Service-Oriented System Engineering (SOSE). IEEE, 2019.
		
		\bibitem{zilliqa_limitations}
		Limitations of Zilliqa’s sharding approach. Available online at: \url{https://medium.com/nearprotocol/limitations-of-zilliqas-sharding-approach-8f9efae0ce3b}
		
		\bibitem{Nightshade}
		Skidanov, Alex, and Illia Polosukhin. "Nightshade: Near protocol sharding design." Available online at: \url{https://nearprotocol. com/downloads/Nightshade. pdf (2019): 39.}
		
		\bibitem{Harmony_protocol}
		Team, Harmony. "Harmony: Technical Whitepaper." (2018).
		
		\bibitem{OmniLedger}
		Kokoris-Kogias, Eleftherios, et al. "Omniledger: A secure, scale-out, decentralized ledger via sharding." 2018 IEEE Symposium on Security and Privacy (SP). IEEE, 2018.
		
		\bibitem{Rapidchain}
		Zamani, Mahdi, Mahnush Movahedi, and Mariana Raykova. "Rapidchain: Scaling blockchain via full sharding." Proceedings of the 2018 ACM SIGSAC Conference on Computer and Communications Security. 2018.
		
		\bibitem{Polkadot}
		Wood, Gavin. "Polkadot: Vision for a heterogeneous multi-chain framework." White paper 21.2327 (2016): 4662.
		
		\bibitem{Cosmos}
		Kwon, Jae, and Ethan Buchman. "Cosmos whitepaper." A Netw. Distrib. Ledgers (2019).
		
		\bibitem{Ethereum_2_0_New_Sharding_2024}
		Danksharding Ethereum 2.0.
		The original link: \url{https://ethereum.org/en/roadmap/danksharding}.
		The backup link saved in the Wayback Machine - Internet Archive website:
		\url{https://web.archive.org/web/20240128121338/https://ethereum.org/en/roadmap/danksharding#what-is-sharding}
		
		\bibitem{Ethereum_number_of_shards}
		Vitalik’s Annotated Ethereum 2.0 Spec.
		The document was written in July-Aug 2020.
		The original link: \url{https://notes.ethereum.org/@vbuterin/SkeyEI3xv#Vitalik\%E2\%80\%99s-Annotated-Ethereum-20-Spec}.
		The backup link saved in the Wayback Machine - Internet Archive website:
		\url{https://web.archive.org/web/20231219093907/https://notes.ethereum.org/@vbuterin/SkeyEI3xv#Vitalik\%E2\%80\%99s-Annotated-Ethereum-20-Spec}
		
		\bibitem{Linear_Scalability_Sharding}
		Mosakheil, Jamal Hayat. "Security threats classification in blockchains." (2018).
		
		\bibitem{beacon_chain_verification}
		Cassez, Franck, Joanne Fuller, and Aditya Asgaonkar. "Formal verification of the ethereum 2.0 beacon chain." Tools and Algorithms for the Construction and Analysis of Systems: 28th International Conference, TACAS 2022, Held as Part of the European Joint Conferences on Theory and Practice of Software, ETAPS 2022, Munich, Germany, April 2–7, 2022, Proceedings, Part I. Cham: Springer International Publishing, 2022.
		
		\bibitem{Merge_blocks}
		Vitalek Buterin, "Merge blocks and synchronous cross-shard state execution." Available online at: \url{https://ethresear.ch/t/merge-blocks-and-synchronous-cross-shard-state-execution/1240}
		
		\bibitem{Synchronous_cross_shard}
		Casey Detrio, "Synchronous cross-shard transactions with consolidated concurrency control and consensus (or how I rediscovered Chain Fibers)" Available online at: \url{https://ethresear.ch/t/synchronous-cross-shard-transactions-with-consolidated-concurrency-control-and-consensus-or-how-i-rediscovered-chain-fibers/2318}
		
		\bibitem{Chainweb}
		Martino, Will, Monica Quaintance, and Stuart Popejoy. "Chainweb: A proof-of-work parallel-chain architecture for massive throughput." Chainweb whitepaper 19 (2018).
		
		\bibitem{Zamfir}
		Vlad Zamfir, Ethereum Sharding Proof of Concept. Available online at: \url{https://github.com/smarx/ethshardingpoc/tree/a0ec249f3fec61279fcde30b403cefebfb23580d#ethereum-sharding-proof-of-concept}
		
		\bibitem{parity}
		"Parity Ethereum client." Available online at: \url{https://github.com/openethereum/parity-ethereum}
		
		\bibitem{Polkadot_Wiki}
		Polkadot Wiki. Available online at: \url{https://wiki.polkadot.network/docs/getting-started}
		
		\bibitem{eth_block_time}
		Ethereum 2.0 Block Time. Available online at: \url{https://github.com/ethereum/consensus-specs/blob/676e216/specs/phase0/beacon-chain.md#time-parameters}
		
		\bibitem{BC4T}
		O’Brien, Dermot, et al. "Final Report of the Exploratory Research Project, Blockchain for Transport (BC4T)." (2022).
		
		\bibitem{Dynamic}
		Tennakoon, Deepal, and Vincent Gramoli. "Dynamic blockchain sharding." 5th International Symposium on Foundations and Applications of Blockchain 2022 (FAB 2022). Schloss Dagstuhl-Leibniz-Zentrum für Informatik, 2022.
		
		\bibitem{Gasper}
		Buterin, Vitalik, et al. "Combining GHOST and casper." arXiv preprint arXiv:2003.03052 (2020).
		
		\bibitem{PoS}
		King, Sunny, and Scott Nadal. "Peercoin: Peer-to-peer crypto-currency with proof-of-stake." self-published paper, August 19.1 (2012).
		
		\bibitem{Casper_FFG}
		Buterin, Vitalik, and Virgil Griffith. "Casper the friendly finality gadget." arXiv preprint arXiv:1710.09437 (2017).
		
		\bibitem{Gasper_developers}
		Ethereum Developers Docs Consensus Mechanism Gasper. Available online at: \url{https://ethereum.org/en/developers/docs/consensus-mechanisms/pos/gasper/}
		
		\bibitem{GHOST}
		Sompolinsky, Yonatan, and Aviv Zohar. "Secure high-rate transaction processing in bitcoin." Financial Cryptography and Data Security: 19th International Conference, FC 2015, San Juan, Puerto Rico, January 26-30, 2015, Revised Selected Papers 19. Springer Berlin Heidelberg, 2015.
		
		\bibitem{Ethereum_2_Phase_0_Github}
		Ethereum 2.0 Phase 0 -- Beacon Chain Fork Choice. Available online at: \url{https://github.com/ethereum/annotated-spec/blob/master/phase0/fork-choice.md}
		
		\bibitem{Polkadot_overview}
		Burdges, Jeff, et al. "Overview of polkadot and its design considerations." arXiv preprint arXiv:2005.13456 (2020).
		
		\bibitem{sybil}
		Douceur, John R. "The sybil attack." International workshop on peer-to-peer systems. Springer, Berlin, Heidelberg, 2002.
		
		\bibitem{honest_or_bust}
		Syta, Ewa, et al. "Keeping authorities" honest or bust" with decentralized witness cosigning." 2016 IEEE Symposium on Security and Privacy (SP). Ieee, 2016.
		
		\bibitem{enhancing_bitcoin_security}
		Kogias, Eleftherios Kokoris, et al. "Enhancing bitcoin security and performance with strong consistency via collective signing." 25th usenix security symposium (usenix security 16). 2016.
		
		\bibitem{Elastico}
		Luu, Loi, et al. "A secure sharding protocol for open blockchains." Proceedings of the 2016 ACM SIGSAC Conference on Computer and Communications Security. 2016.
		
		\bibitem{Sharper}
		Amiri, Mohammad Javad, Divyakant Agrawal, and Amr El Abbadi. "Sharper: Sharding permissioned blockchains over network clusters." Proceedings of the 2021 international conference on management of data. 2021.
		
		\bibitem{spontaneous_sharding}
		Ren, Zhijie, et al. "A scale-out blockchain for value transfer with spontaneous sharding." 2018 Crypto Valley Conference on Blockchain Technology (CVCBT). IEEE, 2018.
		
		\bibitem{Towards_sharding}
		Dang, Hung, et al. "Towards scaling blockchain systems via sharding." Proceedings of the 2019 international conference on management of data. 2019.
		
		\bibitem{Hybrids_on_steroids}
		Behl, Johannes, Tobias Distler, and Rüdiger Kapitza. "Hybrids on steroids: SGX-based high performance BFT." Proceedings of the Twelfth European Conference on Computer Systems. 2017.
		
		\bibitem{Attested_append_only_memory}
		Chun, Byung-Gon, et al. "Attested append-only memory: Making adversaries stick to their word." ACM SIGOPS Operating Systems Review 41.6 (2007): 189-204.
		
		\bibitem{TrInc}
		Levin, Dave, et al. "TrInc: Small Trusted Hardware for Large Distributed Systems." NSDI. Vol. 9. 2009.
		
		\bibitem{MongoDB}
		MongoDB's official documentation. Available online at:
		\url{https://docs.mongodb.com/}
		
		\bibitem{HBase}
		Apache HBase Reference Guide. Available online at:
		\url{https://hbase.apache.org/apache_hbase_reference_guide.pdf}
		
		\bibitem{Bigtable_Documentation}
		Google Cloud Bigtable Documentation. Available online at:
		\url{https://cloud.google.com/bigtable/docs}
		
		\bibitem{Zookeeper}
		Hunt, Patrick, et al. "{ZooKeeper}: Wait-free coordination for internet-scale systems." 2010 USENIX Annual Technical Conference (USENIX ATC 10). 2010.
		Available online at: \url{https://zookeeper.apache.org/}
		
		\bibitem{Zab}
		Junqueira, Flavio P., Benjamin C. Reed, and Marco Serafini. "Zab: High-performance broadcast for primary-backup systems." 2011 IEEE/IFIP 41st International Conference on Dependable Systems \& Networks (DSN). IEEE, 2011.
		
		\bibitem{Riak_handbook}
		Meyer, Mathias. Riak handbook. Technical report, Basho, 2011.
		
		\bibitem{Dynamo}
		DeCandia, Giuseppe, et al. "Dynamo: Amazon's highly available key-value store." ACM SIGOPS operating systems review 41.6 (2007): 205-220.
		
		\bibitem{Couchbase_Documentation}
		Couchbase Documentation. Available online at:
		\url{https://docs.couchbase.com/home/index.html}
		
		\bibitem{Couchbase_analytics}
		Hubail, Murtadha AI, et al. "Couchbase analytics: NoETL for scalable NoSQL data analysis." Proceedings of the VLDB Endowment 12.12 (2019): 2275-2286.
		
		\bibitem{Cassandra_book_1}
		Carpenter, Jeff, and Eben Hewitt. Cassandra: The Definitive Guide,(Revised). " O'Reilly Media, Inc.", 2022.
		
		\bibitem{Cassandra_book_2}
		Strickland, Robbie. Cassandra high availability. Packt Publishing Ltd, 2014.
		
		\bibitem{Cassandra_documentation}
		The official Apache Cassandra documentation. \url{https://cassandra.apache.org/doc/}
		
		\bibitem{Cassandra_paper}
		Lakshman, Avinash, and Prashant Malik. "Cassandra: a decentralized structured storage system." ACM SIGOPS operating systems review 44.2 (2010): 35-40.
		
		\bibitem{DynamoDB}
		Amazon DynamoDB. Available online at \url{https://aws.amazon.com/dynamodb/}
		
		\bibitem{DynamoDB_book_1}
		Amazon DynamoDB Developer Guide.
		ISBN-10 : 9888408771.
		ISBN-13 : 978-9888408771.
		Author: Amazon Web Services.
		(June 26, 2018).
		
		\bibitem{DynamoDB_book_2}
		Vyas, Uchit, and Prabhakaran Kuppusamy. DynamoDB Applied Design Patterns. Packt Publishing Ltd, 2014.
		
		\bibitem{DynamoDB_book_3}
		Deshpande, Tanmay. Mastering DynamoDB. Packt Publishing Ltd, 2014.
		
		\bibitem{Google_Cloud_Whitepapers}
		Google Cloud Whitepapers.
		\url{https://cloud.google.com/whitepapers}
		
		\bibitem{Google_Spanner_original_paper}
		Corbett, James C., et al. "Spanner: Google’s globally distributed database." ACM Transactions on Computer Systems (TOCS) 31.3 (2013): 1-22.
		
		\bibitem{Google_Cloud_Spanner_Documentation}
		Google Cloud Spanner Documentation. Available online at:
		\url{https://cloud.google.com/spanner/docs}
		
		\bibitem{Spanner_Becoming_a_SQL_system}
		Bacon, David F., et al. "Spanner: Becoming a SQL system." Proceedings of the 2017 ACM International Conference on Management of Data. 2017.
		
		\bibitem{ScyllaDB_Documentation}
		ScyllaDB Documentation. Available online at:
		\url{https://docs.scylladb.com/}
		
		\bibitem{ScyllaDB_GitHub_Repository}
		GitHub Repository: ScyllaDB. Available online at:
		\url{https://github.com/scylladb/scylladb/wiki}
		
	\end{thebibliography}
\end{document}